\documentclass[aps,nofootinbib]{revtex4}
\usepackage[usenames]{color}
\usepackage{longtable}
\usepackage{graphicx}
\usepackage{color}

\newcommand{\comments}[1]{}

\begin{document} 

\title{Second-order amplitudes in loop quantum gravity}

\author{\small \em Davide Mamone, Carlo Rovelli}
\affiliation{\small\it Centre de Physique Th\'eorique de Luminy\footnote{Unit\'e mixte de recherche (UMR 6207) du CNRS et des Universit\'es de Provence (Aix-Marseille I), de la M\'editerran\'ee (Aix-Marseille II) et du Sud (Toulon-Var); laboratoire affili\'e \`a la FRUMAM (FR 2291).}, Case 907, F-13288 Marseille, EU} 
\date{\small\today} 

\begin{abstract}

\noindent We explore some second-order amplitudes in loop quantum gravity.   In particular,  we compute some second-order contributions to diagonal components of the graviton propagator in the large distance limit, using the old version of the Barrett-Crane vertex amplitude. We illustrate the geometry associated to these terms.  We find some peculiar phenomena in the large distance behavior of these amplitudes, related with the geometry of the generalized triangulations dual to the Feynman graphs of the corresponding group field theory. In particular, we point out a possible further difficulty with the old Barrett-Crane vertex: it appears to lead to flatness instead of Ricci-flatness, at least in some situations. The observation raises the question whether this difficulty remains with the new version of the vertex. 
\end{abstract}
\maketitle

\section{Introduction}

The problem of quantum gravity is --in  a sense-- a double problem.  First, to find the appropriate theory, which we expected to be a background-independent quantum field theory.  Second, to learn how to extract physics from such a background-independent quantum field theory. This second problem is highly non-trivial, because most, if not all,  of the conventional tools for extracting physics from a quantum field theory rely heavily on the existence of an external metric background. 

A tentative solution to this second problem has been developed in the last years \cite{scattering,carlo,carlo2} and is based on two ingredients.  The first is the boundary formalism \cite{oeckl,book}, which we briefly summarize below.  The second is the idea of computing transition amplitudes order by order in a background-independent expansion, where the order is given by the number of interaction--vertices, or, equivalently, in the number of $n$-simplices of the associated dual cellular complex.  If the theory is expressed as a group field theory (GFT) \cite{GFT}, this expansion amounts to a perturbative expansion in the GFT coupling constant $\lambda$ \cite{book}.  We denote this expansion the ``vertex expansion", and we discuss below its physical viability. 

With only a few exceptions \cite{carlo2,perini}, so far most of the literature has been concentrated on first-order terms in this expansion \cite{simone,SE,alesci,alesci2,Alesci:2008ff}. Here we explore the structure, geometry and physical meaning of some second order terms.\footnote{For an analysis of higher order corrections to the quantum gravity propagator in 3 dimensions see \cite{simone,SE}.}  The importance of studying higher-order terms is multifold. First, it is not yet clear which is the physical regime where the vertex expansion is good; the easiest way to address the problem is to compare the first-order terms with higher ones.  Second, the structure of the expansion is still far from being fully settled: there are open question concerning the correct normalization of the amplitudes, and similar. Again, we think that the best way of addressing these issues is concretely, by studying the terms of the expansion. 

The different terms in the vertex expansion that we study have a simple geometrical interpretation.  Roughly speaking, the lowest order term can be viewed as describing an approximation to General Relativity where the geometry of the spacetime region under consideration is approximated by a single 4-simplex, of variable shape and size. 
Higher-order terms give then approximations where the geometry of the region is approximated by a larger number of glued 4-simplices, each of varying shape and size. Here we illustrate in detail the geometry of some cellular complexes contributing to the second-order approximation, obtained by gluing 4-simplices in this way.  

We restrict ourselves to computing the \emph{diagonal} part of the propagator, instead of writing its full tensorial structure \cite{alesci}. Also, we use the dynamics defined by the old Barrett-Crane vertex \cite{BC}; in particular, the specific model we use is the theory GFT/B (see \cite{book}),  introduced in \cite{gftb,PR2}; the result extends immediately also to the theory GFT/C, introduced in \cite{gftc}, which is characterized by particularly good finiteness properties \cite{finiteness1, finiteness2, finiteness3}. For this reason, the results presented here are a bit out-of-date: they need to be extended to the new-vertex models introduced recently \cite{pereira,Livine:2007vk,flipped,Freidel:2007py,Pereira:2007nh,engle}, which have far better properties.  Nothing seems to prevent such extension, and the work done here should open the way for analyzing the theory more of interest.  Similarly, the semiclassical behavior considered here needs to be compared with the results on the semiclassical behavior of these new models \cite{Barrett:2009gg, Conrady:2009px}. 

We find a certain number of features of the amplitudes, which we summarize in the conclusion section.  Of particular interest is the fact that the amplitude appear to be suppressed, at least in some cases, unless the triangulation admits a \emph{flat} metric.  This is not what we expect for the classical limit, which should be dominated by \emph{Ricci} flatness. This problem can be a further sign of the difficulties of the old Barrett-Crane model: we think that it needs to be seriously addressed in the context of the new models.  

\section{Preliminaries}

We briefly recall the basis of the formalism that we use. This is not a self sufficient introduction: we refer the reader to \cite{carlo2} for complete definitions and details, and to 
\cite{book} for a general introduction.  On the other hand, we address here and we offer some clarification on some general questions that have been raised concerning the approach. 

\subsection{The boundary formalism}

The key idea for extracting physics from a background-independent formulation of quantum field theory  is to compute transition amplitudes associated with a \emph{finite} spacetime region, as functions of the quantum state on the boundary $\Sigma$ of the region \cite{scattering,book,oeckl}.  In particular, the boundary state will include the quantum state of the of the gravitational field, namely the quantum state of the boundary \emph{geometry}.   Physically, this means that we are describing a region of quantum spacetime, as it is observed by apparatuses that take measurements at its boundary --- the key point being that these measurements include (quantum) measurements of distances.  

If we do so, the information about the background geometry of the region is provided \emph{dynamically} by the (measured) boundary quantum-state itself. Formally, if the boundary geometry determines a classical solution of the Einstein's equations in the bulk, then we expect the Feynman integral in the region to be dominated by configurations around the classical one.  In this way, the \emph{interior} background geometry is determined by the \emph{boundary} quantum state: this allows us to define background dependent quantities in the context of a fully background independent bulk theory. 

The main tool of this approach is the boundary functional, formally defined by the functional integral over all fields $\phi$ in the interior region, at fixed boundary value $\varphi$ 
\begin{equation}
W[\varphi] = \int_{\phi|_{{}_{\Sigma}}=\varphi}{D\phi\  {e^{iS[\phi]}}}. 
\label{bfunctional}
	\end{equation}
In a background independent theory, where measure $D\phi$ and action $S[\phi]$ are diff-invariant,  this quantity does not depend on the spacetime location of $\Sigma$.  By contracting this quantity with a state $\Psi[\varphi]$, we obtain a probability amplitude associated to this state
\begin{equation}
\langle W|\Psi \rangle \equiv \int D\varphi \ W[\Sigma]\ \Psi[\phi].
\label{integfunctional}
\end{equation}
This amplitude can then be compared, say, with the amplitude 
\begin{equation}
\langle W| \varphi(x)\varphi(y) |\Psi \rangle 
\label{integfunctional2}
\end{equation}
where $\varphi(x)$ is a field operator creating a quantum excitation over $\Psi$. The quantity 
\begin{equation}
W(x,y;\Psi)=\langle W| \varphi(x)\varphi(y) |\Psi \rangle, 
\label{3}
\end{equation}
where the boundary state $\Psi$ satisfies $\langle\Psi |\Psi \rangle=1$ and
\begin{equation}
\langle W|\Psi \rangle=1, 
\label{norm}
\end{equation}
gives the probability amplitude for a field's quantum, or a ``particle", to propagate from $x$ to $y$ in the background defined by $\Psi$ (on the meaning of ``particle" in this context, see \cite{daniele}).   

This formalism reduces to the standard quantum mechanical formalism on a flat space, if we take $\Sigma$ to be the union of the two hypersurfaces $t=0$ and $t=T$,  and $\Psi$ to be the element $\Psi_{00}=|0\rangle\otimes\langle 0|$ of the tensor product ${\cal H}_{in} \otimes{\cal H}^*_{out}$ of the initial and final state spaces \cite{testa, conrady, doplicher}.  Then, taking $x=(\vec x, T), y=(\vec y, 0)$,
\begin{equation}
W(x,y;\Psi_{00})=\langle 0| \varphi(\vec x)e^{-iHT}\varphi(\vec y) |0 \rangle
=\langle 0| \varphi(x)\varphi(y) |0 \rangle,
\label{4}
\end{equation}
while the normalization condition (\ref{norm}) is clearly satisfied
\begin{equation}
\langle 0|e^{-iHT}|0 \rangle=1.
\end{equation}
But the formalism remains meaningful in a diffeomorphism invariant context because the positions $y$ and $x$ of the incoming and outgoing particles are well-defined with respect to the boundary geometry specified by $\Psi$. That is, $W(x,y;\Psi)$ is not invariant under a coordinate transformation on $x$ and $y$ alone, but it is invariant under a coordinate transformation acting on $x$, $y$, as well as $\Psi$. The \emph{relative} position of the particle with respect to the boundary geometry is diffeomorphism-invariant and the amplitude is therefore well-defined and physically meaningful. 
 
If the background geometry defined by the state  $\Psi$ is flat, then the quantity (\ref{3}) should reduces to the conventional quantum field theoretical two-point function in the weak field limit. In particular, if we compute this quantity in general relativity for the gravitational field, then (\ref{3}) should reduce to the weak-field graviton propagator 
\begin{equation}
W_{\mu\nu\rho\sigma}(x,y;\Psi_{00})\to\langle 0|h_{\mu\nu}(x)h_{\rho\sigma}(y) |0 \rangle
\label{4}
\end{equation}
in the large distance limit.
The dictionary of the translation between the 3-geometry to 3-geometry transition-amplitude language, and the graviton scattering language, is studied in detail in \cite{Mattei:2005cm}.

\subsection{LQG implementation}

The boundary formalism described above can be made concrete in the context of loop quantum gravity (LQG) \cite{lqg,ash}. We take the state $\Psi$ to live in the LQG state space, and we take the boundary functional given by the spinfoam formalism \cite{book}. The compatibility between the canonical loop-theory and the covariant spinfoam-theory, still unclear when the boundary formalism was developing, has since been firmly established \cite{engle}. In particular, a spinfoam dynamics $W$ can be generated by group-field-theoretical methods, where it reproduces the spinfoam model amplitude on any given two-complex, at each order of a vertex expansion \cite{book}.  By choosing an $s$-knot basis $|s\rangle$ in the LQG state space \cite{book}, we have then 
\begin{equation}
W^{abcd}(x,y,q)={\sum_{ss'} W[s]\ \langle s'| h^{ab}(x)h^{cd}(y)|s\rangle\ \Psi_q[s]}. 
\label{propag_LQG}
\end{equation}  
for a state satisfying the normalization condition
\begin{equation}
{\sum_{s} W[s]\ \Psi_q[s]}=1. 
\label{norm_LQG}
\end{equation}  
Concretely, a boundary functional $W[s]$ is defined by any spinfoam model. In fact, a spinfoam model is precisely an algorithm that compute an amplitude $W[s]$ for each boundary spin network $s$. It has the intuitive interpretation as a regularization of the Misner-Wheeler integral-over-4geometries $g$
\begin{equation}
W[q]=\int_{g|_{{}_{\Sigma}}=q} Dg\ e^{iS_{EH}[g]}
\label{norm_LQG}
\end{equation}  
intrinsically regularized by the discreteness of the geometry as established by LQG. Here $S_{EH}[g]$ is the Einstein-Hilbert action and $q$ is the boundary 3-geometry. 

One possible triangulation independent way to write $W[s]$ is to use GFT \cite{GFT}. Here we use the Group Field Theory B (GFT-B) spinfoam model 
\begin{equation}
W[s]=\int D\phi \ f_s(\phi)\ e^{-\int\phi^2+\frac{{\lambda}}{5!}\phi^5}. 
\label{Ws}
\end{equation}
The field $\phi$ is a function on $SO(4)^4$; $s$ is an $s$-knot \cite{book} (or, loosely speaking, a ``spin network") with $n$ nodes and $f_s$ is a polynomial of order $n$ in the fields, obtained contracting the indices of the field following the path  defined by the $s$-knot.  See \cite{book} for the details and the notation. The choice of this spinfoam model here is dictated only by simplicity and convenience. In fact, the limitations of this model are well known, and the results here need to be extended to the more realistic models. 

The Feynman rules of this theory are as follows. The field $\phi$ decomposes in modes $\phi^{\alpha_n}_{j_ni}$ with $n=1,2,3,4$.  Here $j_n$ are $SU(2)$ representations, $\alpha_n$ is an index in the ${j_n}$ representation space, and $i$ labels the (elements of a basis in the space of the) intertwiners in the tensor product of the four representation $j_n$.  The standard quantum field theoretical perturbation expansion in $\lambda$ of this amplitude generates a sum over Feynman graphs with five-valent vertices.    The propagator is 
\begin{equation}
{P}_{\alpha_{n}i}^{j^s_{n}}\,{}_{\alpha'_{n}i'}^{j^s_{n}}=
\sum_s
\mathcal{P}_{\alpha_{n}i}^{j^s_{n}}\,{}_{s(\alpha'_{n})i'}^{s(j^s_{n})}
\label{s}
\end{equation}
where the sum is over all permutations $s$ of four elements and 
\begin{equation}
\mathcal{P}_{\alpha_{n}i}^{j^s_{n}}\,{}_{\alpha'_{n}i'}^{j^s_{n}}=
\delta_{ii'} \prod_n \delta_{\alpha_{n}\alpha'_n}\delta^{j^s_{n},j_n'}.
\end{equation}
The vertex is five-valent, and is given by 
\begin{equation}
\mathcal{V}_{\alpha_{nm}i_n}^{j_{nm}}=
\lambda
\mathcal{B}(j_{nm})
\prod_{n\ne m} \delta_{\alpha_{nm}\alpha_{mn}}\delta^{j_{nm},j_{mn}}
\end{equation}
where $\mathcal{B}(j_{nm})$ is the vertex amplitude; see \cite{book}. For large spins,
the asymptotic behavior of the vertex amplitude is given by 
\begin{equation}
\mathcal{B}(j_{nm}) \sim e^{i S_{\rm Regge}(j_{nm})}+e^{-i S_{\rm Regge}(j_{nm})}+ 
D(j_{nm})
\end{equation}
where $S_{\rm Regge}(j_{nm})$ is the Regge action associated to a 4-simplex with areas proportionals to $j_{nm}$ and $D(j_{nm})$ is a factor that appears when the areas $j_{nm}$ define degenerate configurations of the 4-simplex. The Regge action has the form
\begin{equation}
S_{\rm Regge}(j_{nm})=\sum_{mn}\phi_{nm}(j_{nm})\ j_{nm}
\end{equation}
where $\phi_{nm}(j_{nm})$ are the dihedral angles of the 4 simplices with areas proportional to the $j_{nm}$. 

Each choice of a permutation $s$ at every propagator determines a pattern of contractions of the $\delta_{\alpha\alpha'}$ delta functions; a closed set of contractions
($\delta_{\alpha_1\alpha_2}\delta_{\alpha_2\alpha_3}...\delta_{\alpha_n\alpha_1}$) determines a sequence of propagators, and is called a ``face". Thus the sum (\ref{s}) becomes a sum over the two-complexes with four-valent edges having the Feynman graph as 1-skeleton.  In particular, if the two-complex is dual to a 4d triangulation, then the associated amplitude can be shown to be a Feynman sum for a discretization of general relativity on that Regge-like triangulation \cite{flipped,engle}.  Thus, the group field theory generates a discretized Feynman sum for general relativity, where the sum is extended over --appropriately generalized-- triangulations \cite{book}.  This is why it can be seen as a discretizations of the Misner-Hawking sum-over-4geometries.

\subsection{The vertex expansion}

The GFT formulation suggests a perturbative expansion for $W[s]$: the expansion in the GFT coupling constant $\lambda$, namely the vertex expansion. The physical meaning of this expansion is clarified by noticing that individual terms of this expansion can be equally obtained by truncating general relativity to the finite-number-of-degrees-of-freedom system formed by a Regge triangulation on a given discretization of spacetime.  Notice that is neither a short-scale, nor a large-scale expansion, since the individual 4-simplices can be large or small  \cite{flipped}. It is rather more similar to the approximation used very effectively in cosmology, where only some degrees of freedom of the geometry of the universe are left free \cite{cosmo}.  

Can such a truncation provide an interesting approximation to the quantum gravitational dynamics?  The truncation is background-independent, in the sense in which Regge calculus is. But one may worry that on a fixed triangulation the theory has a finite number of degrees of freedom and therefore it cannot sufficiently capture the field-like behavior of gravity.  This objection is wrong, since it would apply to the standard perturbative expansion of QED as well: if we compute a scattering process between a finite number $m$ of particles to a finite order $n$ in perturbative QED, we are restricting the QED Fock space to the subspace formed by a finite number of particle (as many as $n$ vertices can produce from $m$ particles).  Thus we are \emph{de facto} truncating QCD to a theory with a finite number of degrees of freedom (a particle has obviously a finite number of degrees of freedom).  In other words, conventional QFT perturbation expansion \emph{includes} a truncation of the field theory to a theory with a finite number of degrees of freedom. There is no reason for the same not be viable in gravity.   

The correct question, then, is not if the truncation given by the vertex expansion yields a viable approximation, but rather in which regime this approximation is viable.  Again, the Regge-lattice analogy provides the answer: any gravitational physics that can be captured by the finite Regge triangulation. For instance if a phenomenon is characterized by a size (wavelength) $l$ and can be confined in a region of size $L$, then $L/l$ sets the scale of the relevant number of ``cells" needed to approximate the phenomenon.   

Lattice QCD provides a good example of this: effective lattice QCD calculation yield the correct mass spectrum of the hadrons using lattices that have a rather small number of cells. This number is determined by the ratio between the size of the hadron and the minimal relevant wavelength.  What is remarkable is that good quantum physics is obtained with cubic lattices with sides of only a few cells.  Clearly there is no really need of infinite lattices to do physics.

\subsection{The large distance expansion}

We are interested in the graviton two-point function (\ref{3}) at first order in $\lambda$,
in the limit in which the boundary geometry is large.  In this limit we are only looking at very large wavelengths, and it is therefore reasonable to expect that the vertex expansion is viable.   The calculation of the graviton two-point function (\ref{3}) in this limit at first order in $\lambda$ on the basis of the formalism described above was completed in \cite{carlo} for the diagonal terms ($\mu=\nu$ and $\rho=\sigma$), and in \cite{alesci} for the other terms.  The fact that the non-diagonal terms of the propagator turned out to be wrong was a main reason for the replacement of the old version of the Barrett-Crane vertex with its new version \cite{pereira,Freidel:2007py,flipped,engle}.  The correct $1/L^2$ dependence of the propagator on the distance $L$, obtained in \cite{carlo}, was confirmed in a next to leading order evaluation \cite{carlo2}.  Several second-order terms are considered below. 

\newpage

\section{Joining two 4-simplices}

\subsection{One internal propagator:  $4\rightarrow1\rightarrow4$ Pachner's move}

Consider the second-order Feynman diagram 
\begin{center}
  \includegraphics[height=2cm]{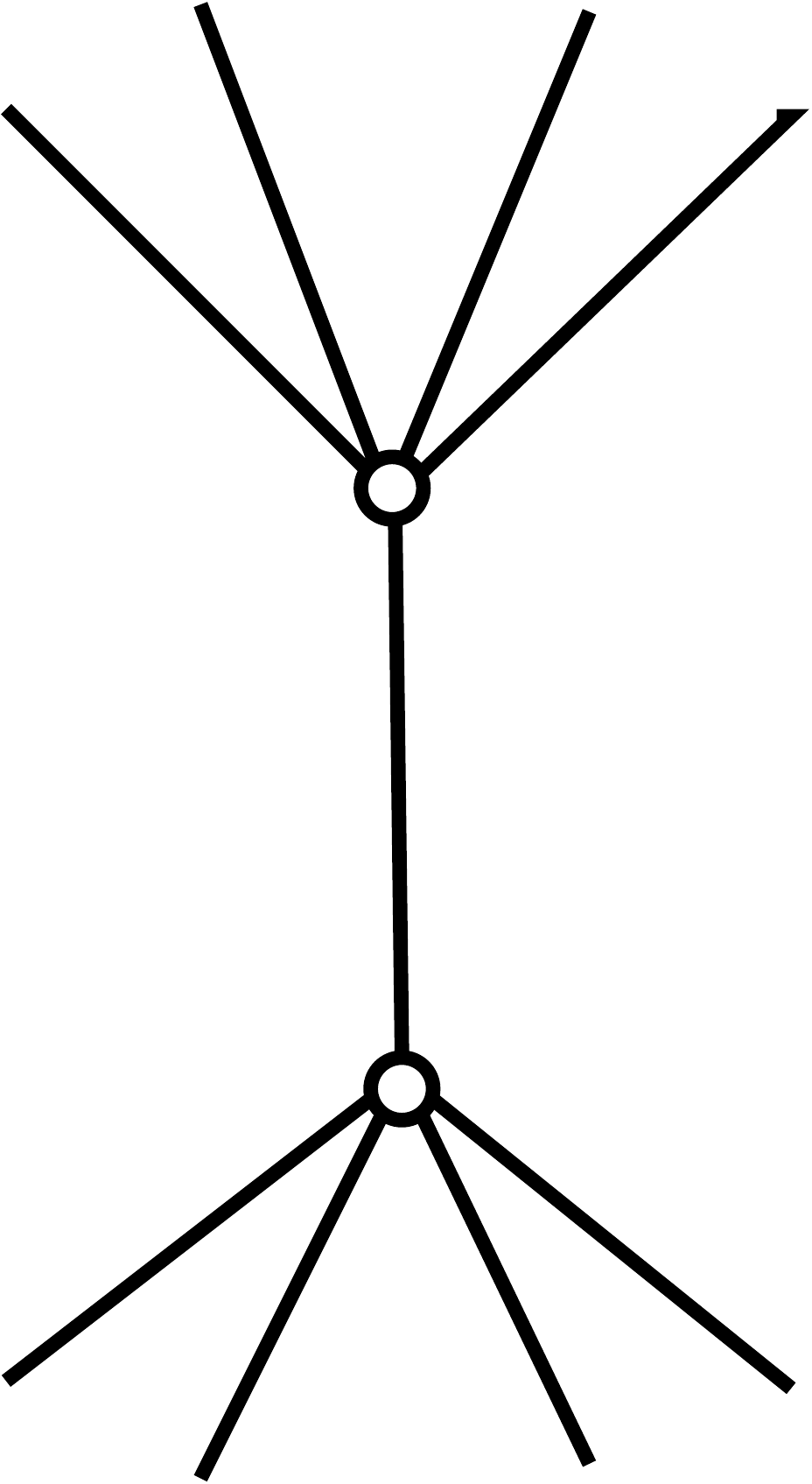}\hspace{1em}. 
  \end{center}
Call  $v^{\rm u}$ and $v^{\rm d}$ (for $up$ and $down$) the two vertices of this graph, $e$ the internal propagator, and $e^{\rm u}_{n}$ and $e^{\rm d}_{n}$, where $n=1,2,3,4$ the external legs. 

\begin{figure}[h]
\begin{center}
\includegraphics[height=6cm]{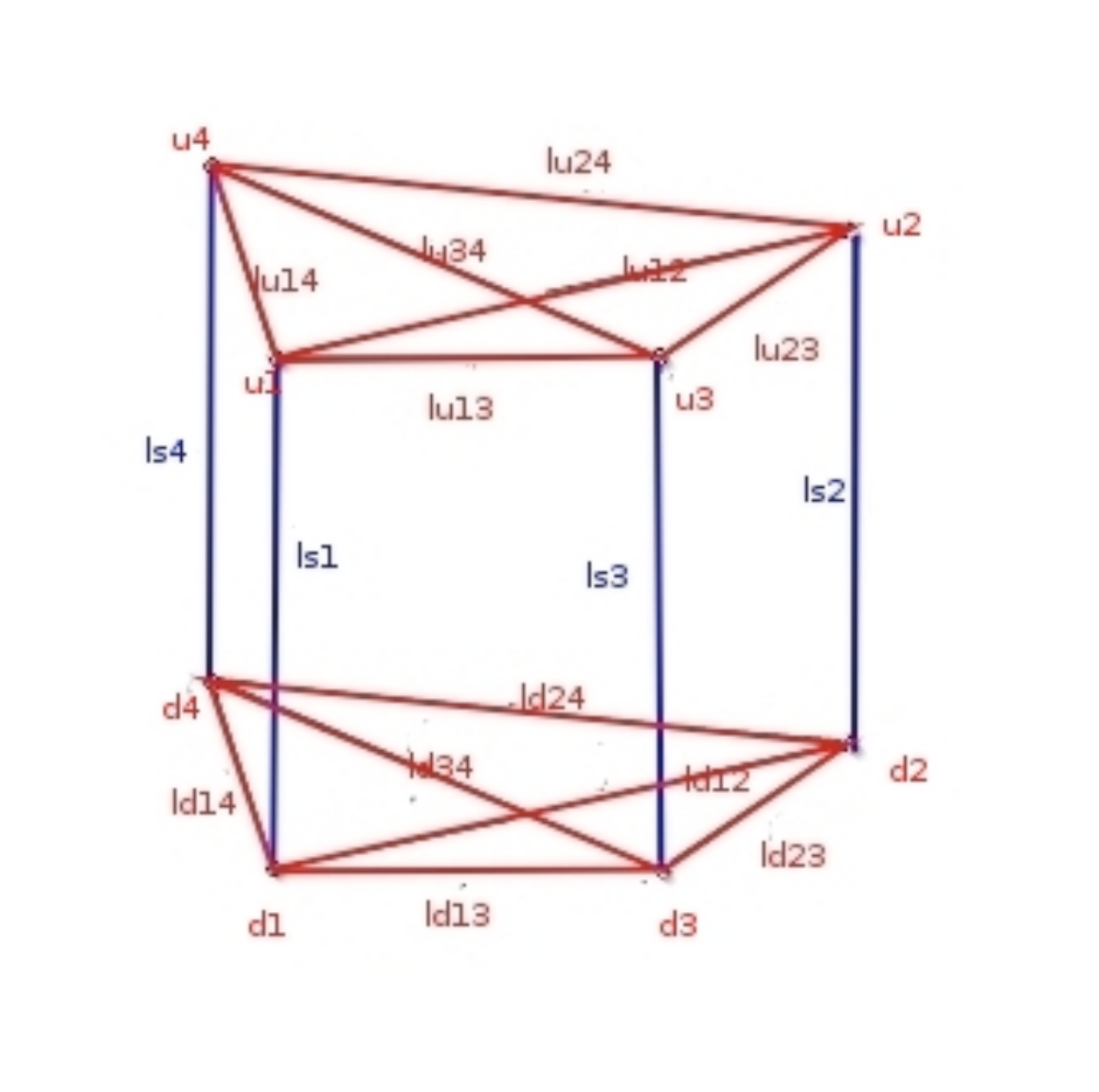}
\end{center}\vskip-6mm
\caption{\label{Boundarys_8}  The boundary spin network $s_8$.}
\end{figure}

Such a graph can appear in computing the amplitude associated to the observable 
$f_{s_8}(\phi)$ determined by the spin network $s_8$  illustrated in Figure \ref{Boundarys_8}. The graph of this spin network is $\Gamma_8$.  It consists of two tetrahedral spin networks connected by four links. The spin network  $s_8$ is obtained by coloring the links and nodes of $\Gamma_8$.  We denote the spins and intertwiners associated with nodes and links of  $\Gamma_8$ as in Figure \ref{Boundarys_8}, Panel b. That is, we denote $j^{\rm u}_{nm},j^{\rm d}_{nm},j^s_{n}$ the twenty representations associated with the 20 links $l^{\rm u}_{nm},l^{\rm d}_{nm}, l^s_{n}$ of $\Gamma_8$, and $i_n^{\rm u}, i_n^{\rm d}$ the eight intertwiners associated with the eight nodes $n^{\rm u}_{n}$ and $n^{\rm d}_{n}$.   The set $s_8=(\Gamma_8, j^{\rm u}_{nm},j^{\rm d}_{nm},j^s_{n}, i^{\rm d}_{n}, i^{\rm u}_{n})$ defines a boundary spin network.

The observable $f_{s_8}(\phi)$ determined by this spin network is  
\begin{equation} 
f_{s_8}(\phi) = \sum_{\alpha_{nm}\beta_{nm}}\prod_{n=1,4} 
\phi^{\alpha_{nm}  i^{\rm u}_n}
_{j^{\rm u}_{nm}}\ 
\phi^{\beta_{nm}  i^{\rm d}_n}
_{j^{\rm d}_{nm}}
\end{equation}
where we have used the notation $j^{\rm u}_{nn}:=j^{\rm d}_{nn}:=j^s_{n}$
This is a monomial of order eight in the field. The expansion of its expectation value
at order $\lambda^2$ gives
\begin{equation}
W[s_8] = \frac{\lambda^2}{2 (5!)^2} \int \mathrm{D} \phi \, f_{s_8}(\phi) \, \left(\int \phi^5 \right)^2 \, \mbox{e}^{-\int \phi^2}
\end{equation}
\begin{figure}[b]
\begin{center} 
\includegraphics[scale=0.40]{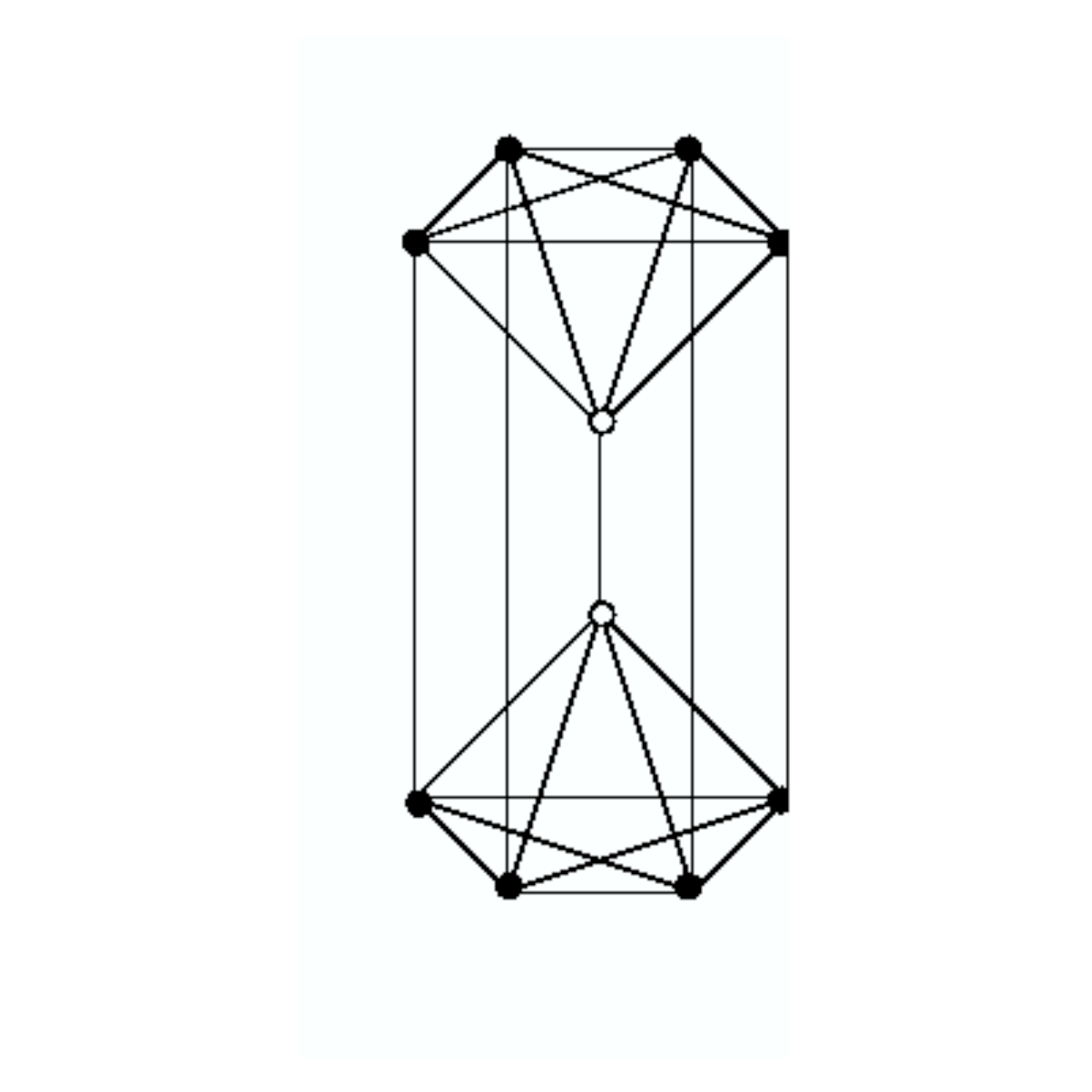}
\vskip-8mm
\caption{
\label{4} Feynman graph and boundary spinnetwork}\label{dual1}
\end{center} 
\end{figure}

The Wick expansion of this integral gives two vertices and nine propagators.
(If $d$ is the order of $\phi$ in $f_s$, and $n_v$ is the number of vertices, then the number of propagators is clearly $n_p=\frac{d+5n_v}{2}$). In particular, consider the
term where the four $up$ (resp. $down$) legs of the graph are connected to the upper (resp. lower) tetrahedral spin network, as in Figure \ref{4}; the corresponding amplitude is  
\begin{equation}
\label{ws_8}
W[s_8]=
\begin{tiny}
\Big(\prod_{n=1,4}{P}_{\alpha_{nm}\alpha_{n}i^{\rm u}_n}^{j^{\rm u}_{nm}\,j_n}\,{}_{\alpha'_{nm}i'_n}^{j'_{nm}}\Big)
\mathcal{V}_{\alpha'_{nm}\gamma_mi'_ni''}^{j'_{nm}j''_m}
{P}_{\gamma_mi''}^{j''_m}\,{}_{\delta_mi'''}^{j'''_m}
\mathcal{V}_{\delta_{m}i'''\beta'_{nm}i'_n}^{j'''_{m}j''_{nm}}\Big(\prod_{n=1,4}
{P}_{\beta_{nm}\beta_{n}i^{\rm d}_n}^{j^{\rm d}_{nm}\,j_n}\,{}_{\beta'_{nm}i'_n}^{j''_{nm}}\Big)\end{tiny}.
\end{equation} 
In the first parenthesis we have the contribution from the $up$ boundary (contractions among  the ``up" four boundary tetrahedra $u_n$); then we have the contraction between these and the fifth (taken as ``internal"); then the  internal propagator; then the down vertex and the contractions in the `down" boundary.

The sums over permutations (\ref{s}) in the propagators give rises to a sum over 
two-complexes having the Feynman graph as two-skeleton.   Since we are interested in the large $j$ behavior of the amplitude, and since each face of the two-complex carries some powers of $j$, the dominant term will be the one with the maximum number of faces.  It is not hard to see that this term is given by the term
\hspace{-3cm}
\begin{eqnarray}
\label{ws_8}
W[s_8]&=&
\begin{tiny}
\Big(\prod_{n=1,4}\mathcal{P}_{\alpha_{nm}\alpha_{n}i^{\rm u}_n}^{j^{\rm u}_{nm}\,j_n}\,{}_{\alpha'_{nm}i'_n}^{j'_{nm}}\Big)
\mathcal{V}_{\alpha'_{nm}\gamma_mi'_ni''}^{j'_{nm}j''_m}\mathcal{P}_{\gamma_mi''}^{j''_m}\,{}_{\delta_mi'''}^{j'''_m}
\mathcal{V}_{\delta_{m}i'''\beta'_{nm}i'_n}^{j'''_{m}j''_{nm}}\Big(\prod_{n=1,4}
\mathcal{P}_{\beta_{nm}\beta_{n}i^{\rm d}_n}^{j^{\rm d}_{nm}\,j_n}\,{}_{\beta'_{nm}i'_n}^{j''_{nm}}\Big)
\end{tiny}	\nonumber  
\\ &=&
\begin{tiny}
 \Big(\prod_{n=1,4}
\dim(i^{\rm u}_{n})\dim(i^{\rm d}_{n})\dim(j_n)\Big)
\Big(\prod_{n<m, n=1,4}
\dim(j^{\rm u}_{mn})\dim(j^{\rm d}_{mn})\Big)
\mathcal{B}(j_{nm})\mathcal{B}(j_{nm})
\end{tiny}
\label{path}
\end{eqnarray} 
of the sum over permutations.  Let us analyze this term. It is obtained by 
adding sixteen faces to the Feynman graph: four faces $f_{n}$ bounded by the three edges $e^{\rm u}_{n}, e, e^{\rm d}_{n}$, six faces $f^{\rm u}_{nm}$ bounded by the two edges $e^{\rm u}_{n}$ and $e^{\rm u}_{m}$ and six faces $f^{\rm d}_{nm}$ bounded by the two edges $e^{\rm d}_{n}$ and $e^{\rm d}_{m}$.   The nodes $n^{\rm u}_{n}$ and $n^{\rm d}_{n}$ of $\Gamma_8$ bound the edges $e^{\rm u}_{n}$ and $e^{\rm d}_{n}$ respectively.  The links $l^s_{n}, l^{\rm u}_{nm}$ and $l^{\rm d}_{nm}$ of this graph bound the faces  $f_{n}, f^{\rm u}_{nm}$ and $f^{\rm d}_{nm}$, respectively. 

\begin{figure}[h]
\begin{center}
\includegraphics[height=5cm]{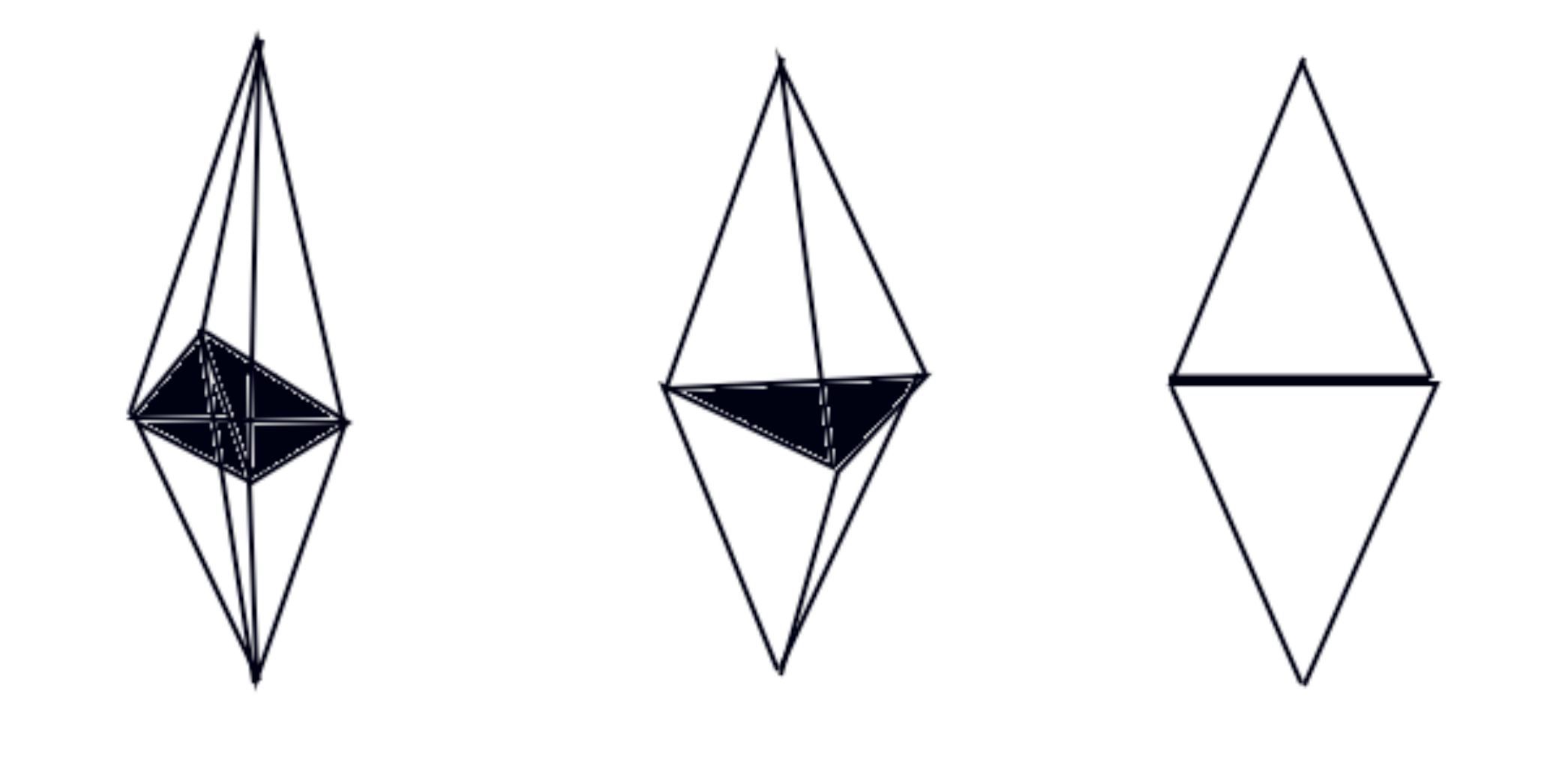}
\end{center}
\caption{\label{3dcreaz} The spacetime triangulation $\Delta_8$ and the 3d and 2d analogs.}
\end{figure}
The structure of this two-complex becomes transparent by noticing that it is the complex dual to a rather simple triangulation, which we call $\Delta_8$. This is obtained by gluing two 4-simplices by one tetrahedron  $\mathcal{T}$.  The four--dimensional triangulation $\Delta_8$ is illustrated in the first panel of  Fig.\,\ref{3dcreaz}.  This is the 4d analog of the 3d and 2d cases illustrated in the other two panels of the figure.  

The triangulation $\Delta_8$ is formed by 6 points, which we label as $1,2,3,4,u,d$; by the 14 edges ${(n,m),(n,u),(n,d)}$; the 12 faces $(n,m,n),(n,m,u),(n,m,d)$; the 9 tetrahedra  $(1,2,3,4),(n,m,p,u),(n,m,p,d)$ and the two 4-simplices  $(1,2,3,4,u),(1,2,3,4,d)$.  Here $n=1,2,3,4$ and $n\ne m \ne p$. 

The two vertices $v^{\rm u}$ and $v^{\rm d}$ are dual to the two 4-simplices of this triangulation. The four triangles that bound  $\mathcal{T}$ are the dual to the faces $f_{n}$; the other six triangles of the upper (resp. lower) 4-simplex are dual to the faces $f^{\rm u}_{nm}$ (resp. $f^{\rm d}_{nm}$).  Notice that all these triangles belong to the boundary of the triangulation, as is clear from the 3d analog.  The graph dual to this boundary is clearly the $\Gamma_8$ graph of Figure \ref{Boundarys_8}:  the four upper nodes of $\Gamma_8$ correspond to the four upper tetrahedra; the four lower nodes of $\Gamma$ correspond to the four lower tetrahedra. The six links joining the upper (lower) nodes correspond to the six upper (lower) vertical triangles $t^{\rm u}_{nm}$ ($t^{\rm d}_{nm}$); the four vertical links correspond to the four triangles $t_{n}$ bounding $\mathcal{T}$.  Therefore 
$\Gamma_8$ is the dual of a triangulation of a compact 3d surface, with the topology of a three-sphere, which can be viewed as the boundary of the spacetime region formed by 
two adjacent 4-simplices.

\begin{figure}[h]
\begin{center} 
\makebox{\includegraphics[height=8 cm]{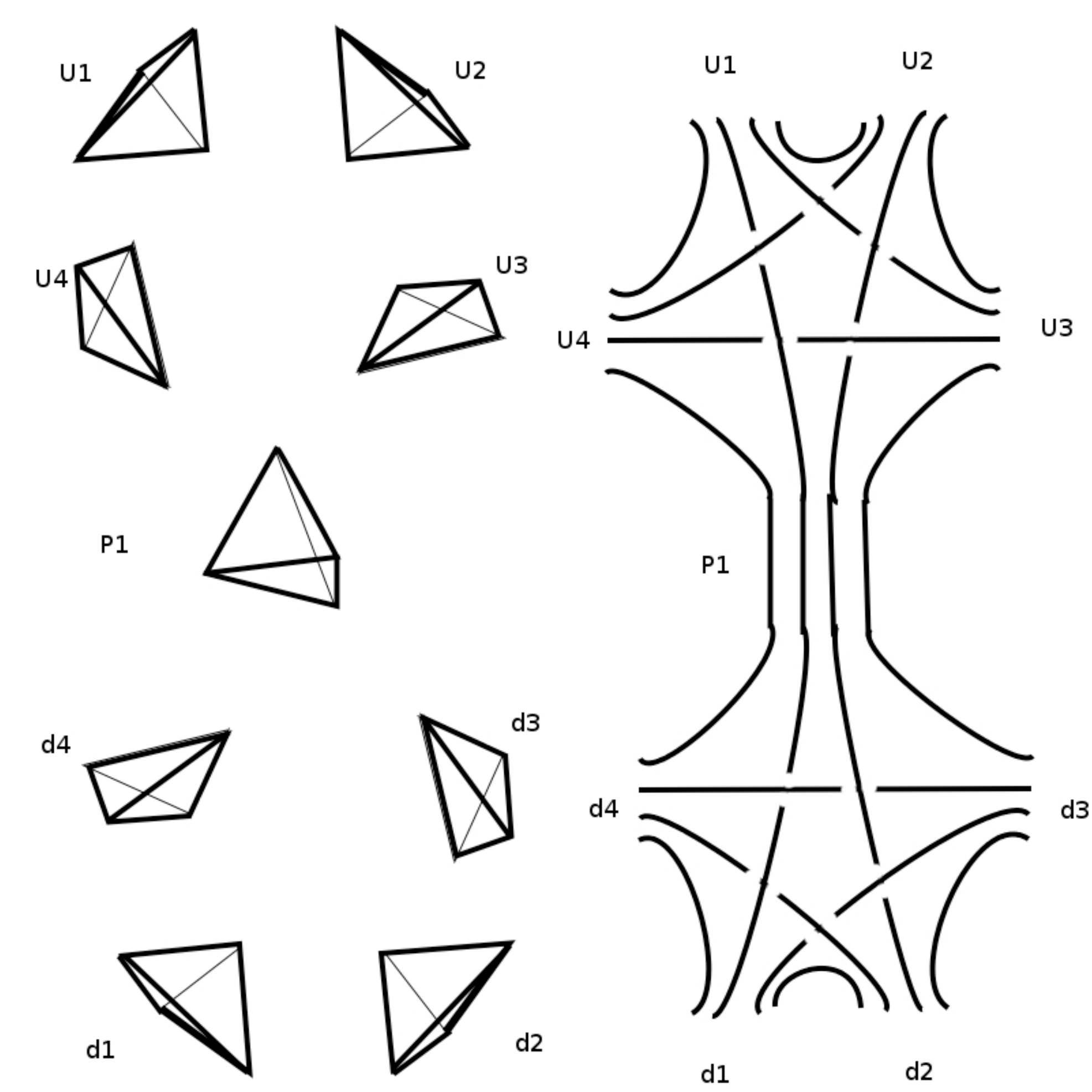}}
\ \ \raisebox{4cm}{\begin{small}
\begin{tabular}{ccccc}
& & & & \\ \hline
\multicolumn{1}{|c|}{Tetrahedra} & \multicolumn{4}{|c|}{Triangles} \\
\hline
& & & &  \\
\hline
\multicolumn{1}{|c|}{$u_1$}  & \multicolumn{3}{c}{\textcolor{red}{$t^u_{23}$}   \textcolor{red}{$t^u_{24}$} \textcolor{red}{$t^u_{34}$}}  & \multicolumn{1}{c|}{\textcolor{blue}{$t_{234}$}} \\
 \hline
\multicolumn{1}{|c|}{$u_2$} & \multicolumn{3}{c}{\textcolor{red}{$t^u_{13}$}  
\textcolor{red}{$t^u_{14}$} \textcolor{red}{$t^u_{34}$}} & \multicolumn{1}{c|}{\textcolor{blue}{$t_{134}$}} \\ \hline
\multicolumn{1}{|c|}{$u_3$} & \multicolumn{3}{c}{\textcolor{red}{$t^u_{12}$} 
\textcolor{red}{$t^u_{14}$} \textcolor{red}{$t^u_{24}$}} & \multicolumn{1}{c|}{\textcolor{blue}{$t_{124}$}}  \\ \hline
\multicolumn{1}{|c|}{$u_4$}  & \multicolumn{3}{c}{\textcolor{red}{$t^u_{12}$}   \textcolor{red}{$t^u_{13}$} \textcolor{red}{$t^u_{23}$}} &  \multicolumn{1}{c|}{\textcolor{blue}{$t_{123}$}} \\
 \hline
& & & &  \\
& & & &  \\
 
\hline
\multicolumn{1}{|c|}{$p_1$} & \textcolor{blue}{$t_{234}$} & \textcolor{blue}{$t_{134}$} & \textcolor{blue}{$t_{124}$} & \multicolumn{1}{c|}{\textcolor{blue}{$t_{123}$}} \\ \hline
& & & &  \\
& & & &  \\
\hline

\multicolumn{1}{|c|}{$d_1$}  & \multicolumn{3}{c}{\textcolor{red}{$t^d_{23}$}   \textcolor{red}{$t^d_{24}$} \textcolor{red}{$t^d_{34}$}}  & \multicolumn{1}{c|}{\textcolor{blue}{$t_{234}$}} \\
 \hline
\multicolumn{1}{|c|}{$d_2$} & \multicolumn{3}{c}{\textcolor{red}{$t^d_{13}$}  
\textcolor{red}{$t^d_{14}$} \textcolor{red}{$t^d_{34}$}} & \multicolumn{1}{c|}{\textcolor{blue}{$t_{134}$}} \\ \hline
\multicolumn{1}{|c|}{$d_3$} & \multicolumn{3}{c}{\textcolor{red}{$t^d_{12}$} 
\textcolor{red}{$t^d_{14}$} \textcolor{red}{$t^d_{24}$}} & \multicolumn{1}{c|}{\textcolor{blue}{$t_{124}$}}  \\ \hline
\multicolumn{1}{|c|}{$d_4$}  & \multicolumn{3}{c}{\textcolor{red}{$t^d_{12}$}   \textcolor{red}{$t^d_{13}$} \textcolor{red}{$t^d_{23}$}} &  \multicolumn{1}{c|}{\textcolor{blue}{$t_{123}$}} \\
\hline
\end{tabular}
\end{small}}
\caption{\label{dia}
a: {\it Tetrahedral decomposition}: the up 4-simplex (tetrahedra $u_1u_2u_3u_4p_1$) is glued to the down 4-simplex ($p_1d_1d_2d_3d_4$) through the shared tetrahedron $p_1$. b: {\it GFT diagram}: every line corresponds to a triangle, four lines grouped to a tetrahedron. c: Relation between tetrahedra and triangles.}
\end{center}
\end{figure}

The path of the indices in (\ref{path}) gives the geometrical decomposition of the triangulation illustrated in Fig.\ref{dia}, where each line corresponds to a face of the two complex, or, equivalently, a triangle of the triangulation. 

If we interpret the vertical axis as a ``time" axis, the triangulation $\Delta_8$ represents the world-history of a point $d$ opening up to a tetrahedron $\mathcal{T}$ and then recollapsing to a point $u$.   (In the 3d case, we have a point opening up to a triangle and then recollapsing; in the 2d case, we have a point opening up to a segment and then recollapsing. See Fig.\ref{3dcreaz}.) The process described by this amplitude can therefore be interpreted as a creation and annihilation of an ``atom of space" \cite{carlo2}.  

Following \cite{carlo2}, let us now use the amplitude (\ref{path}) for computing a contribution to the graviton two-point function, at given boundary state.  At this order, the relevant component of the boundary state is on the graph $\Gamma_8$
\begin{equation}
\Psi_{\mathbf q}[s_8]=\Psi_{\mathbf q}( j^{\rm u,d}_{nm},j^{\rm s}_{n}). 
\end{equation}
Let us assume that the boundary state describes a regular semiclassical geometry for the boundary of the triangulation $\Delta_8$.  Let this be peaked on the geometry $\mathbf q_8$ defined by the 
the boundary of the region of $R^4$ formed by the two pyramidal 4-simplices having 
for basis a regular tetrahedron $\mathcal T$ with side of length $L'$, and height $T$.  
Since we are interested in the large $j$ regime, we peak the state on the values
\begin{eqnarray}
j^{\rm s(8)}_n&=&   \frac{\sqrt{3}}{32\pi \hbar G }{L'}^2 \equiv j_L,   
\label{jL}
\\
 j^{\rm u(8)}_{nm}=j^{\rm d(8)}_{nm}&=&  \frac{1}{8\pi  \hbar G }   \left(2{L'} \sqrt{\frac{{L'}^2}{8} + T^2} + \frac{\sqrt{3}}{2} {L'}^2\right) \equiv j_{TL}.
\label{jTL}
\end{eqnarray}
We do not give here the explicit value of the background dihedral angles $\Phi_l^{\scriptscriptstyle (8)}$, which can be obtained by elementary geometry: 
for details see appendix in \cite{carlo2}. 
\begin{figure}[h]
\begin{center}
  \includegraphics[height=4cm]{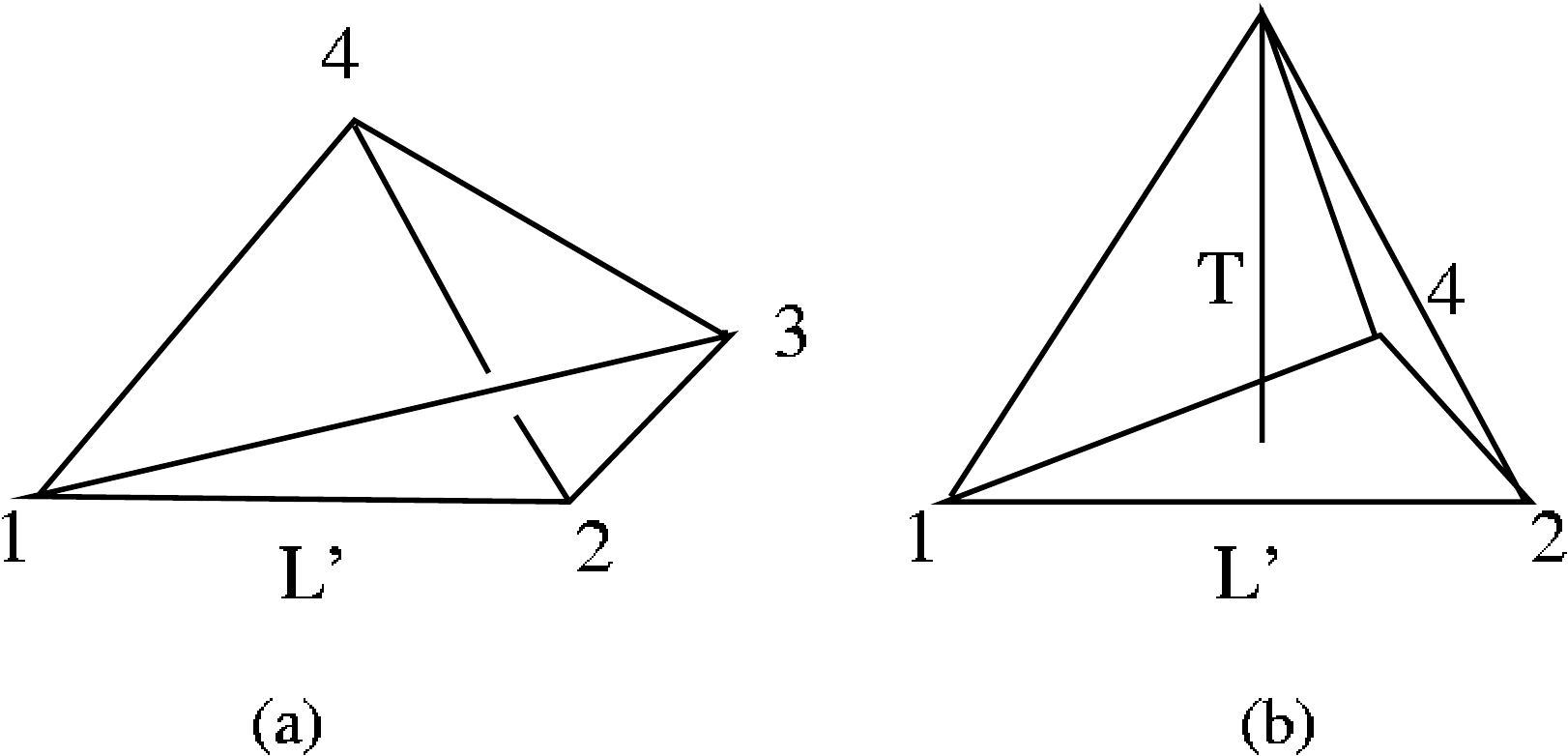}
\caption{\label{Tetra1} Central tetrahedron $(a)$. Lateral tetrahedron $(b)$.}
  \end{center} 
  \end{figure}
We choose a boundary state given by a Gaussian peaked on  $\mathbf q_8$. Writing all spins in a single vector $j_l=(j_{nm}^{\rm u},j_{nm}^{\rm d}, j^{\rm s}_{n})$, we have   
\begin{equation}
\Psi_{\mathbf q}[s_8] = 
C_8 \ e^{
- \alpha_{ll'}{(j_l- j_l^{\scriptscriptstyle (8)})(j_{l'}- j_{l'}^{\scriptscriptstyle (8)})}
+i \Phi_l^{\scriptscriptstyle (8)}j_{l}}. 
\label{vuoto1}
\end{equation} 
Following \cite{carlo,carlo2}, we contract the four indices of (\ref{propag_LQG}) with normals to boundary triangles, and choose in particular, say, to look at the ``diagonal" term determined by the triangles $t_{12}^{\rm u}$ and $t_{13}^{\rm d}$, obtaining
\begin{equation}
{\mathbf G_{s_8}}({L'},T)\equiv (n^{\rm u}_{12})_a(n^{\rm u}_{12})_b(n^{\rm d}_{13})_c(n^{\rm d}_{13})_d W^{abcd}(x,y, {\mathbf q_8})=\sum_{ss'} W[s]\langle s'| \delta j^{\rm u}_{12}\delta j^{\rm d}_{23} |s\rangle\Psi_q[s]
\end{equation} 
where $\delta j = j-j^{(8)}$. All the terms in this expression are now well defined. 

We now use the asymptotic expression for each $\mathcal{B}$, as in \cite{carlo}. This is given by the cosine of the Regge action plus the degenerate term.  The phase in the boundary state suppresses the sum unless it is matched by a corresponding phase of a term in $W[s]$.  This happens for only one of the exponentials in the cosine, as can be seen as follows.   The sum of the Regge actions for the two 4-simplices,
$S_{Regge}=S_{Regge}^{\rm u}+S_{Regge}^{\rm d}$ can be 
expanded around $j_{TL}$ and $j_L$
\begin{equation}
S_{Regge}(j_{nm}^{\rm u},j_{nm}^{\rm d}, j^{\rm s}_n) =
\tilde \phi^{(8)}_{n m} j_{n m}^{\rm u} +
 \tilde \phi^{(8)}_n j^{\rm s}_n + 
\tilde \phi^{(8)}_{n m} j_{n m}^{\rm d} +
 \tilde \phi^{(8)}_n j^{\rm s}_n 
+ \frac{1}{2} G_{ll'} \, \delta j_l \delta j_{l'}, 
\label{exp}
\end{equation}
where $\tilde \phi^{(8)}_n$ and $\tilde \phi^{(8) }_{n m}$ are the dihedral angles of flat 4-simplices with the given boundary \emph{intrinsic} geometry;
the linear terms in the expansion of the Regge action sum up, giving the dihedral angle of the boundary of the 4d region, which is precisely the sum of the dihedral angles 
of the two 4-simplices at the faces of $\cal T$. That is, 
$\tilde\phi^{(8)}_{nm} = \Phi^{(8)}_{nm}$, but   $2 \tilde \phi^{(8)}_{n}=\Phi^{(8)}_{n}$.  
The second order term in (\ref{exp}) is the ``discrete derivative" \cite{carlo2}
\begin{equation}
G_{ll'} = \left(\frac{\delta^2 S_{Regge}}{\delta j_{l} 
\, \delta j _{l'}}\right)_{j_l  = j^{(8)}_l} .
\end{equation}
This matrix can be computed from elementary geometry. Being a derivative of an angle with respect of an area, $G_{ll'}$ should scale as the inverse of $\sqrt{j^{(8)}_l j^{(8)}_{l'}}$. It is therefore convenient to define the scaled quantity 
\begin{equation}
\Gamma_{ll'} = \frac{G_{ll'}}{\sqrt{j^{(8)}_l j^{(8)}_{l'}}}.
\end{equation}

Thus, we obtain
\begin{equation}
{\mathbf G_{s_8}}({L'},T)=\frac{4\lambda^2{\cal N}_8 }{j_{TL}^2} \, 
 \sum_{\delta j_{l}} 
\delta j^{\rm u}_{12} \,  \delta j^{\rm d}_{13}  \, 
P_{\tau}^2 \, e^{-i\left(S_{Regge}^{\rm u} + S_{Regge}^{\rm d} 
+ k_{\tau} \frac{\pi}{2}\right)} \mbox{e}^{
-\tilde\alpha_{ll'}(j_l- j_l^{\scriptscriptstyle (8)})(j_{l'}- j_{l'}^{\scriptscriptstyle (8)})
+ i  \Phi^{(8)} j_l}
\end{equation} 
where ${\cal N}_8$ is fixed by the normalization condition. If only the Feynman graph that we are considering enters it, we have 
\begin{equation}
\frac{1}{{\cal N}_8 }=\frac{4\lambda^2}{j_{TL}^2} \, 
 \sum_{\delta j_{l}} 
\, 
P_{\tau}^2 \, e^{-i\left(S_{Regge}^{\rm u} + S_{Regge}^{\rm d} 
+ k_{\tau} \frac{\pi}{2}\right)} \mbox{e}^{
-\tilde\alpha_{ll'}(j_l- j_l^{\scriptscriptstyle (8)})(j_{l'}- j_{l'}^{\scriptscriptstyle (8)})
+ i  \Phi^{(8)} j_l}
\end{equation} 
The Gaussian peaks the sums around the background values. We can therefore expand the summand around these values.  The first order term of the expansion of the Regge action around these values cancels the phases in the state, leaving
\begin{eqnarray}
 {\mathbf G_{s_8}}(L',T)=\frac{4{\cal N}_8 \lambda^2} {j_{TL}^2}
 \sum_{\delta j_{l}} 
\delta j^{\rm u}_{12} \,  \delta j^{\rm d}_{13} \,
\mbox{e}^{- \frac{1}{2} \delta {\bf j}^T \tilde\mathcal{A} \, \delta{\bf j}},
 \end{eqnarray} 
\noindent where we have introduced the matrix 
\begin{eqnarray}
\tilde\mathcal{A}_{ll'} = 2\tilde\alpha_{ll'} +i  G_{ll'} = \sqrt{j^{(0)}_l j^{(0)}_{l'}}\ (2\alpha +i  \Gamma)_{ll'}  
 = \sqrt{j^{(0)}_l j^{(0)}_{l'}}\ \mathcal{A}_{ll'} 
\label{matrix}
\end{eqnarray}
\noindent and the vector $\delta{\bf j} =(\delta j_l)= (\delta j^{\rm u}, \delta j^{\rm d}, \delta j^{\rm s})$.  Approximating the sum with gaussian integrals gives 
\begin{eqnarray}
{\mathbf G_{s_8}}(L', T) = \frac{16 \pi}{j_{TL}^2}
 \big(\tilde\mathcal{A}\big)^{-1}_{j_{12}^{\rm u}, j_{13}^{\rm d}}= 
\frac{16 \pi}{j_{TL}}
 \big(\mathcal{A}\big)^{-1}_{j_{12}^{\rm u}, j_{13}^{\rm d}}
\end{eqnarray}
which is proportional to $1/j_L$, as in the first order calculation \cite{carlo}.
Thus we recover the expected  $\frac{1}{L^2}$ behavior of the linearized theory.

\subsection{Two internal propagators: $3\rightarrow2 \rightarrow 3$ Pachner's move}

\begin{figure}[h]
\begin{center}
\includegraphics[height=6cm]{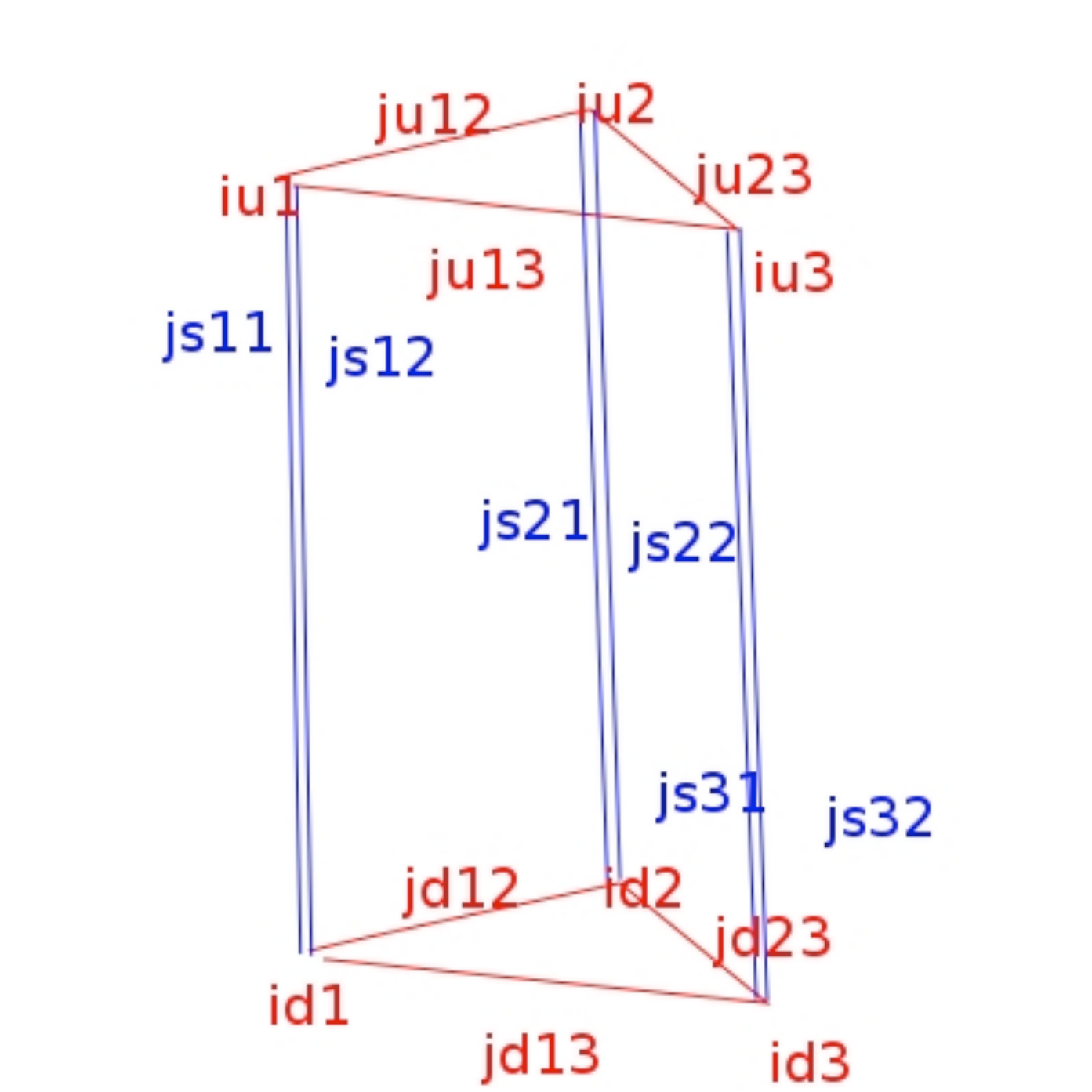}
\caption{The boundary spin network $s_6$.}
\label{spinet2}
\end{center}\vskip-3mm 
\end{figure}

 Consider the Feynman diagram \vskip-3mm
  \begin{center}
  \includegraphics[height=2cm]{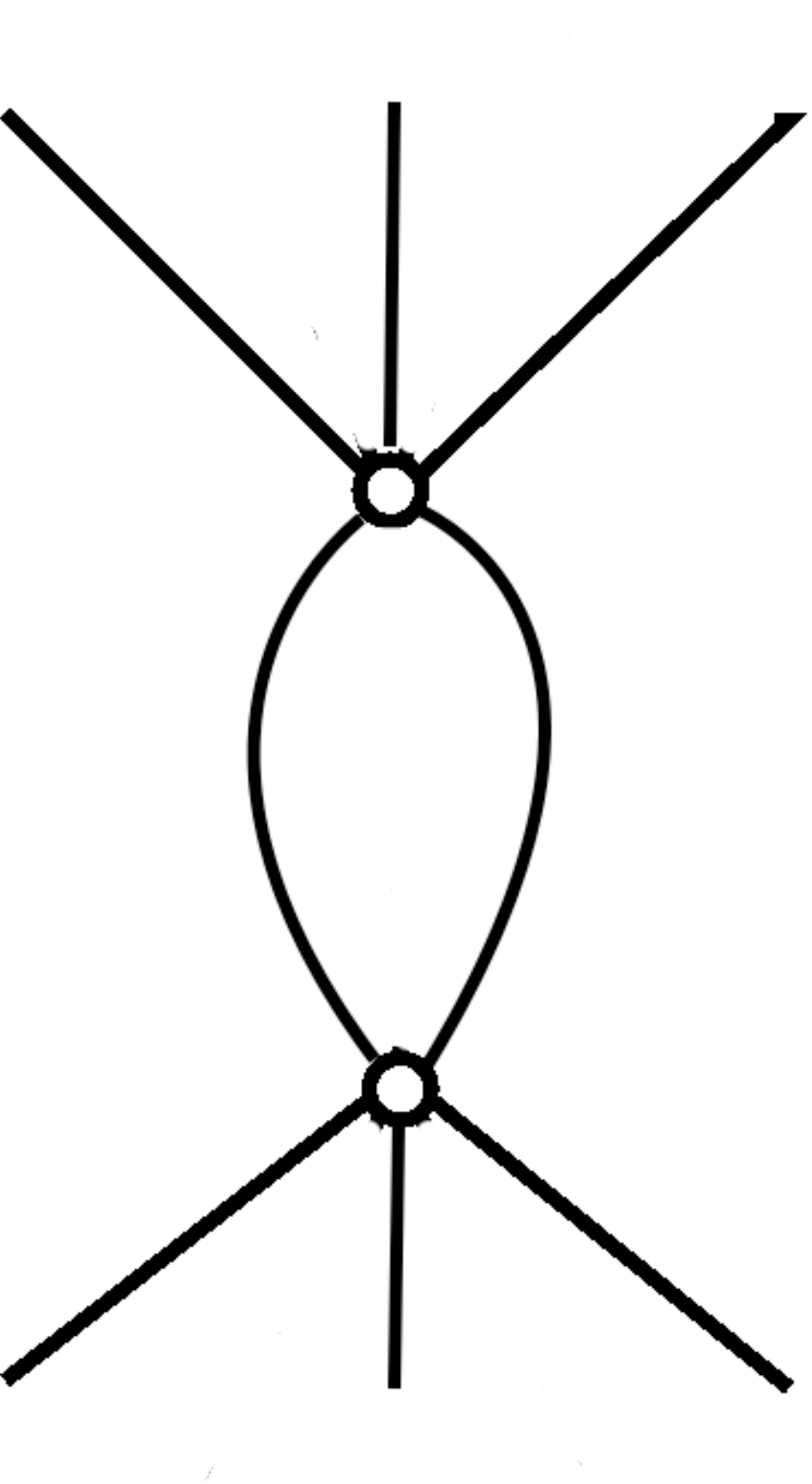}\ \ \ .
  \end{center}
This will appear in the amplitude of an observable $f_{s_6}$ defined by a spin network with graph $\Gamma_6$, illustrated in Fig.\ref{spinet2}, consisting of two triangular
spin networks connected by six links: three ``up"  nodes $u_n$,  and three ``down" nodes $d_n$, $n=1,2,3$.

It is convenient to denote the links of this spin network as follows. Call  $l^{\rm u}_{nm}$ (resp $l^{\rm d}_{nm}$) with $n\ne m$ and $n,m=1,2,3$ the three upper (resp. lower) links, and denote $l^{\rm s}_{nv}$ with $v=4,5$ the two links joining $u_n$ and $d_n$.  Denote $j^{\rm u}_{nm},j^{\rm d}_{nm},j^{\rm s}_{nv}$ the representations
associated to the 12 links $l^{\rm u}_{nm},l^{\rm d}_{nm}, l^{\rm s}_{nv}$ of $\Gamma_6$, and $i_n^{\rm u}, i_n^{\rm d}$ the six 
intertwiners associated to the six nodes $n^{\rm u}_n$ and $n^{\rm d}_n$. 
The set $s_6=(\Gamma_6,  j^{\rm u}_{nm},j^{\rm d}_{nm},j^{\rm s}_{nv} i_n^{\rm u}, i_n^{\rm d})$ is the boundary spin network we consider in this section.  
 
\begin{figure}[b]
\begin{center}
\includegraphics[scale=0.40]{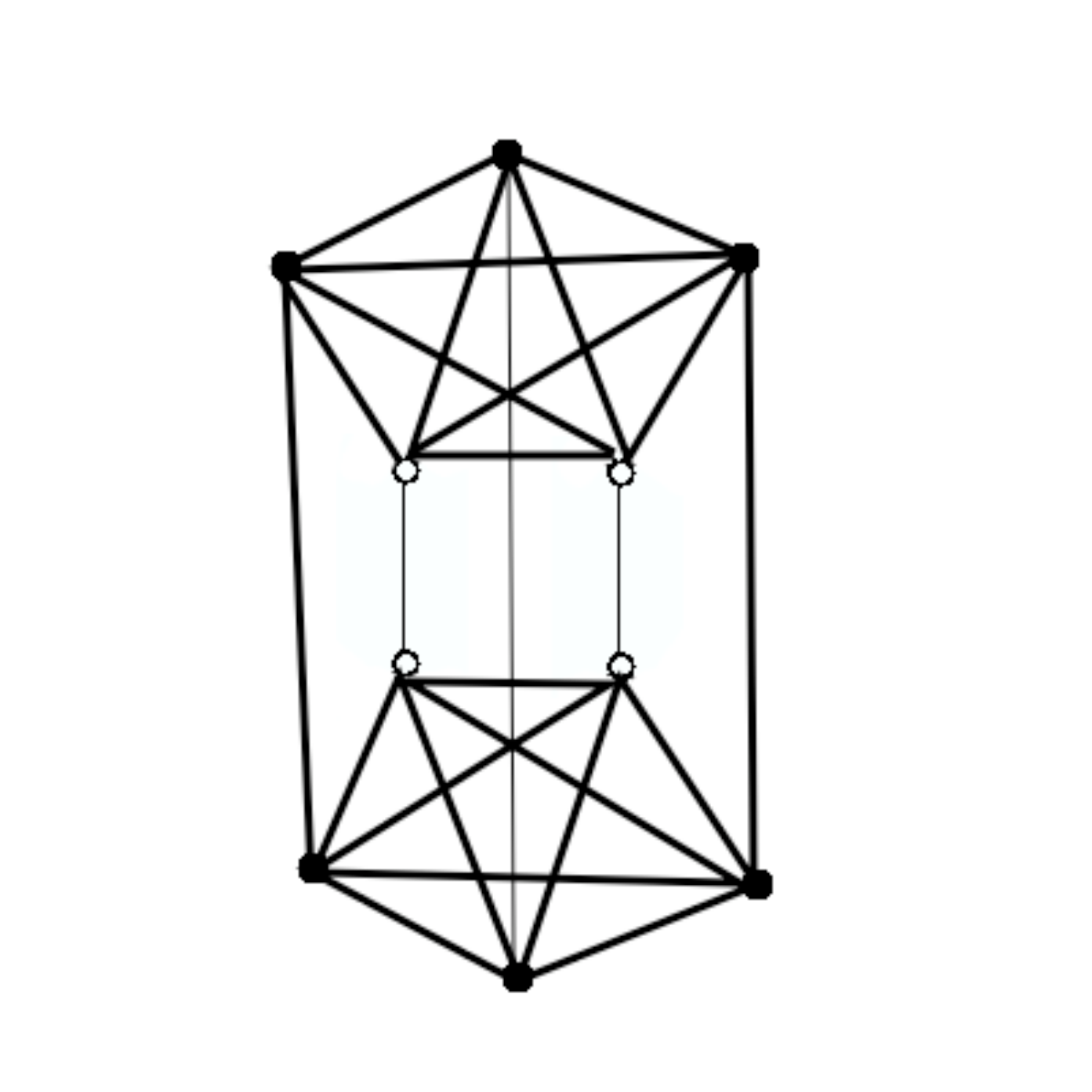}
\caption{Gluing two 4-simplices via two tetrahedra. Dots represent the tetrahedra; lines represent triangles.  The empty dots connected by a thin line are identified. The lines emerging from connected empty dots are identified as well.}
\label{spinet222}
\end{center}
\end{figure}

The boundary function $f_{s_6}(\phi)$ for this spin network is a monomial of order six in the field: 
\begin{equation} 
f_{s_6}(\phi) = \sum_{\{\alpha\}}\prod_{n=1,2,3} 
\phi^{\alpha^{\rm u}_{nm}  i^{\rm u}_n}
_{j^{\rm u}_{nm}j^{\rm s}_{nv}}\ 
\phi^{\alpha^{\rm d}_{nm} i^{\rm d}_n}
_{j^{\rm d}_{nm}j^{\rm s}_{nv}r}
\end{equation}
where $n\ne m=1,...,5$. At order $\lambda^2$, the corresponding amplitude 
\begin{eqnarray}
W[s_6] = \frac{\lambda^2}{2 (5!)^2} \int \mathrm{D} \phi \, f_{s_6}(\phi) \, \left(\int \phi^5 \right)^2 \, 
\mbox{e}^{-\int \phi^2}
\end{eqnarray}
gives two vertices and eight propagators: 
\begin{eqnarray}
W[s_6]&=& 
\begin{tiny}  
\Big(\prod_{n=1,3}\mathcal{P}_{\alpha_{nm}\alpha_{n}i^{\rm u}_n}^{j^{\rm u}_{nm}\,j_n}\,{}_{\alpha'_{nm}i'_n}^{j'_{nm}}\Big)
\mathcal{V}_{\alpha'_{nm}\gamma_m \eta_m i'_n i''i''''}^{j'_{nm}j''_mj''''_{m}}
\mathcal{P}_{\gamma_m i''}^{j''_m}\,{}_{\delta_m i'''}^{j'''_m}
\end{tiny}
\\ 
\nonumber
& &\hspace{2em}\times\ \ \ \ 
\begin{tiny}  
\mathcal{P}_{\eta_m i''''}^{j''''_m}\,{}_{\theta_mi'''''}^{j'''''_m}
\mathcal{V}_{\theta_m i'''''\delta_{m}i'''\beta'_{nm}i'_n}^{j'''''_{m}j'''_{m}j''_{nm}} 
 \Big(\prod_{n=1,3}\mathcal{P}_{\beta_{nm}\beta_{n}i^{\rm d}_n}^{j^{\rm d}_{nm}\,j_n}\,{}_{\beta'_{nm}i'_n}^{j''_{nm}}\Big)
 \end{tiny}
\end{eqnarray} 
As before, the two complex with the highest number of faces is the one with 
$\mathcal{P}$ replaced by $P$. This is dual to the (generalized) triangulation
$\Delta_6$, obtained by gluing two 4-simplices via two tetrahedra. This is 
schematically indicated in Figure \ref{spinet222}.  

The triangulation $\Delta_6$ is formed by 5 points, which we label as $1,2,3,4,5$; by the 11 edges ${(n,m),(n,4),(n,5), (4,5)_u, (4,5)_d}$ (here $n=1,2,3$; notice that there are \emph{two} distinct edges connecting the points 4 and 5); the 13 faces $(1,2,3), (n,m,4),(n,m,5),(n,4,5)_u, (n,4,5)_d$; the 8 tetrahedra  $(1,2,3,4),(1,2,3,5),(n,m,4,5)_u,(n,m,4,5)_d$ and the two 4-simplices  $(1,2,3,4,5)_u,(1,2,3,4,5)_d$. 
 
\begin{figure}[h]
  \begin{center}
  {\includegraphics[height=5cm]{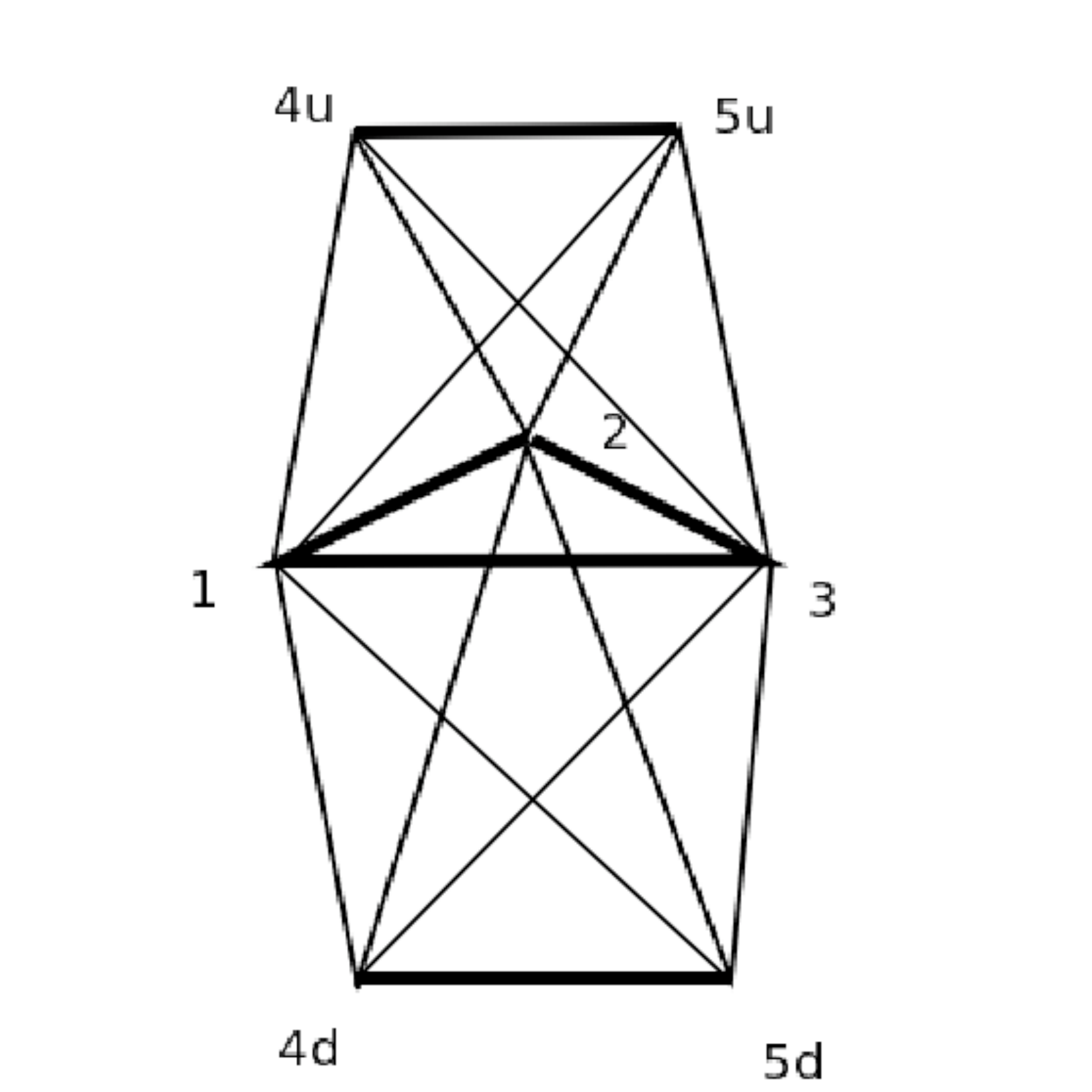}}
  \raisebox{-1cm}{\includegraphics{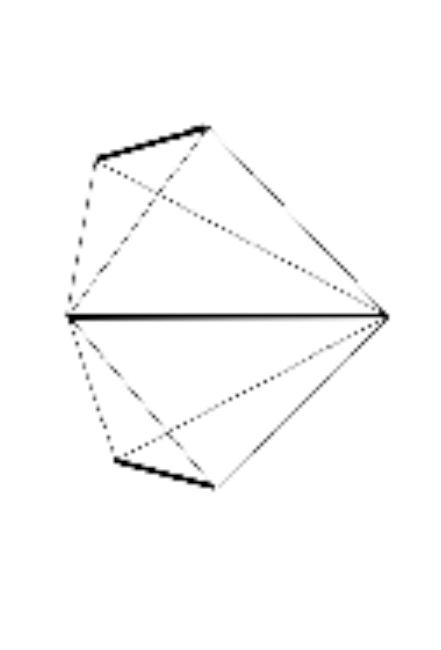}}
  \end{center}
  \vspace{-.8cm}
  \caption{\label{sup_2p} a): The spacetime triangulation $\Delta_6$. The point $5u$ and $5d$ must be identified, and so the points $4u$ and $4d$. The triangles $2-3-5u$ must be identified with the triangle $2-3-5d$, and so on.  The line $4u-5u$ must \emph{not} be identified with the line $4d-5d$.  b): The 3d analog of $\Delta_6$; here as well the the two ends of the upper 
  horizontal line must join the two ends of the lower horizontal line, and the two side triangles of the upper tetrahedron must be identified with the two side triangles of the lower tetrahedron.}
\end{figure}

\begin{figure}[b]
  \begin{center}
  \includegraphics[height=4cm]{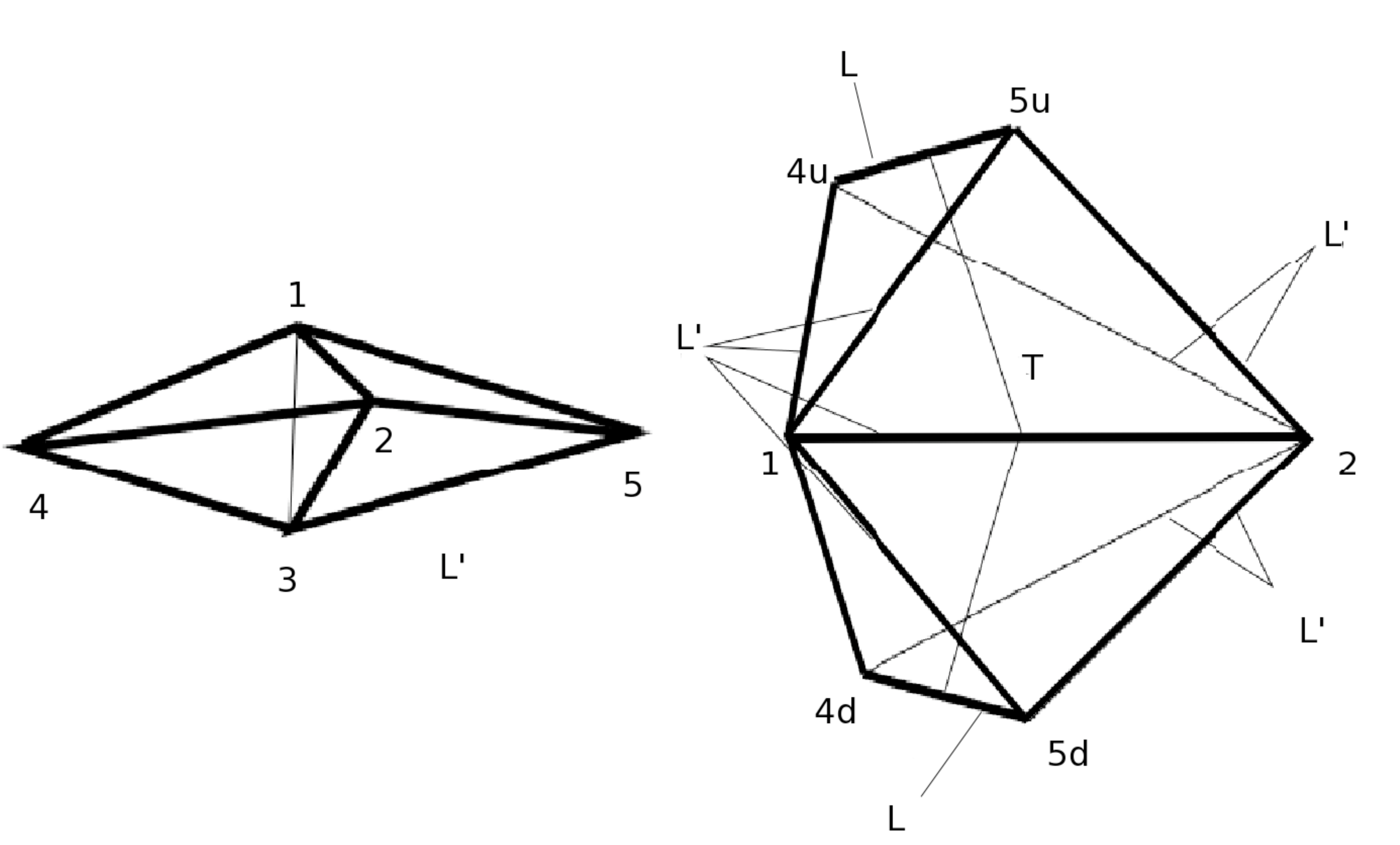}
  \end{center}
  \caption{\label{tetrabordo2} The labeling of the vertices of the central tetrahedra $(a)$ and the lateral tetrahedra $(b)$.}
\end{figure}

We use also the notation $p_4\equiv(1,2,3,4)$ and $p_5\equiv(1,2,3,5)$ for these two tetrahedra, $\mathcal{T}_2=p_4\cup p_5$ their union, and $\tau\equiv(1,2,3)$ the triangle that separates them.  The triangulation can be interpreted as representing the world-history of the line $(4,5)_d$ opening up to the volume $\mathcal{T}_2$, and then recollapsing to the line $(4,5)_u$. The initial and final line join at both ends. See Figure \ref{sup_2p} and Figure \ref{tetrabordo2}. 

\begin{figure}[h]
\begin{center} 
\includegraphics[height=8 cm]{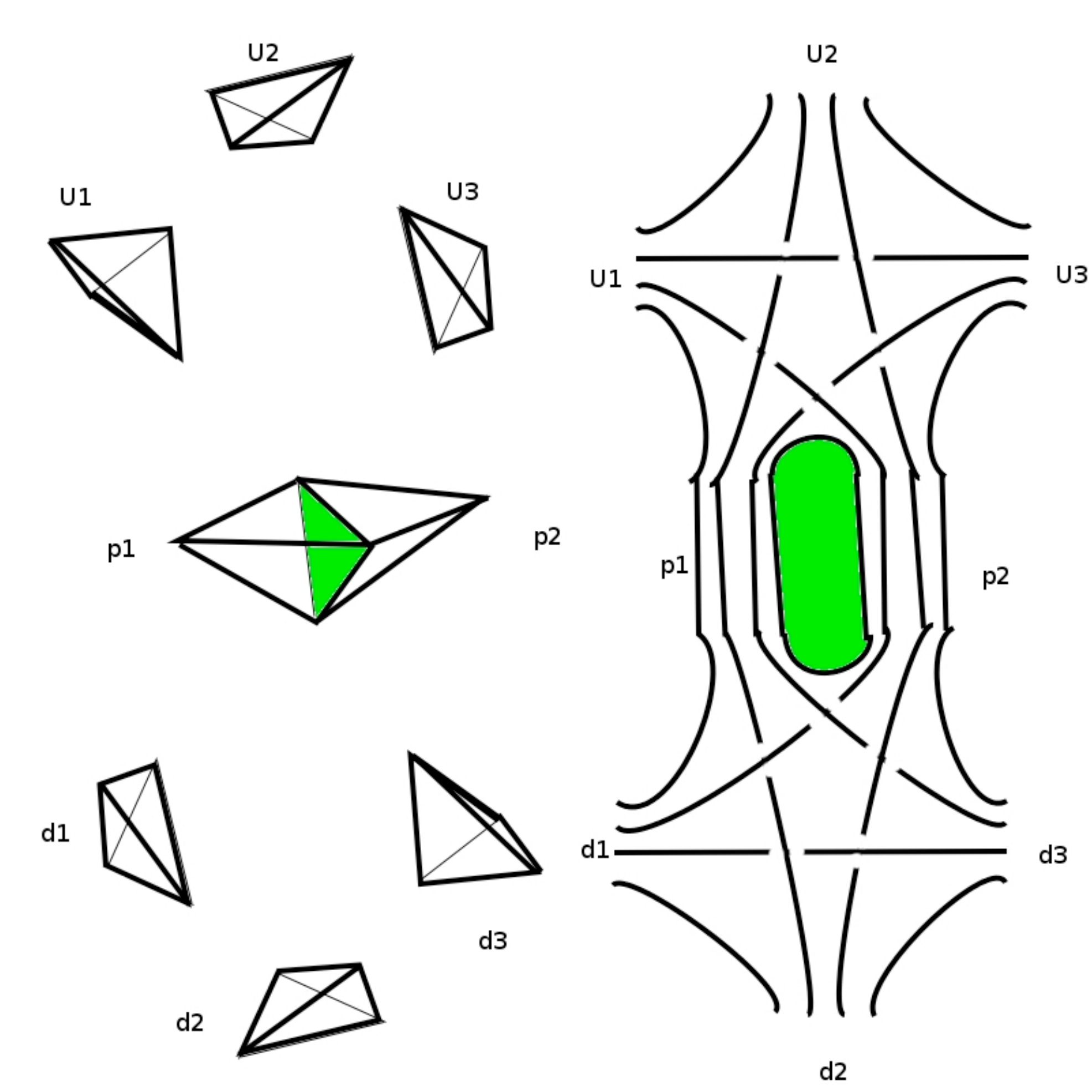}\hspace{2em}
\raisebox{4cm}{\small\begin{tabular}{ccccc}
& & & & \\ \hline
\multicolumn{1}{|c|}{Tetrahedra} & \multicolumn{4}{|c|}{Triangles} \\
\hline
& & & &  \\
\hline
\multicolumn{1}{|c|}{$u_1$}  & \multicolumn{2}{c}{\textcolor{red}
{$t^u_{2}$}   \comments{\textcolor{red}}
{$t^u_{3}$}}  & \multicolumn{2}{c|}{\textcolor{blue}
{$t_{234}$} \comments{\textcolor{blue}}
{$t_{235}$}} \\
 \hline
\multicolumn{1}{|c|}{$u_2$} & \multicolumn{2}{c}{\textcolor{red}
{$t^u_{1}$}  
\comments{\textcolor{red}}
{$t^u_{3}$}} & \multicolumn{2}{c|}{\textcolor{blue}
{$t_{134}$} \comments{\textcolor{blue}}
{$t_{135}$}} \\ \hline
\multicolumn{1}{|c|}{$u_3$} & \multicolumn{2}{c}{\textcolor{red}
{$t^u_{1}$} 
\comments{\textcolor{red}}
{$t^u_{2}$}} & \multicolumn{2}{c|}{\textcolor{blue}
{$t_{124}$} \comments{\textcolor{blue}}
{$t_{125}$}}  \\ \hline
& & & &  \\
& & & &  \\
\hline
\multicolumn{1}{|c|}{$p_1$} & \comments{\textcolor{blue}}
{$t_{234}$} & \comments{\textcolor{blue}}
{$t_{134}$} &
\comments{\textcolor{blue}}
{$t_{124}$} & \multicolumn{1}{c|}{\textcolor{green}
{$t_{123}$}} \\

 \hline
\multicolumn{1}{|c|}{$p_2$} & \comments{\textcolor{blue}}
{$t_{235}$} & \comments{\textcolor{blue}}
{$t_{135}$} &
\comments{\textcolor{blue}}
{$t_{125}$} & \multicolumn{1}{c|}{\textcolor{green}
{$t_{123}$}} \\

\hline
& & & &  \\
& & & &  \\
\hline

\multicolumn{1}{|c|}{$d_1$}  & \multicolumn{2}{c}{\textcolor{red}
{$t^d_2$} \comments{\textcolor{red}}
{$t^d_{3}$}}  &
\multicolumn{2}{c|}{\textcolor{blue}
{$t_{234}$} \comments{\textcolor{blue}}
{$t_{235}$}} \\
 \hline
\multicolumn{1}{|c|}{$d_2$} & \multicolumn{2}{c}{\textcolor{red}
{$t^d_{1}$}  
\comments{\textcolor{red}}
{$t^d_{3}$}} & \multicolumn{2}{c|}{\textcolor{blue}
{$t_{134}$} \comments{\textcolor{blue}}
{$t_{135}$}} \\ \hline
\multicolumn{1}{|c|}{$d_3$} & \multicolumn{2}{c}{\textcolor{red}
{$t^d_{1}$} 
\comments{\textcolor{red}}
{$t^d_{2}$}} & \multicolumn{2}{c|}{\textcolor{blue}
{$t_{124}$} \comments{\textcolor{blue}}
{$t_{125}$}}  \\ \hline
& & & &  \\ 
\end{tabular}}
\caption{{\it Tetrahedral decomposition}: a) two 4-simplices share two 
tetrahedra $p_1$ and $p_2$ \label{diag2}.
b){\it GFT diagram}: 3 up and 3 down boundary lines,
 6 propagators lines and 1 internal loop (green). c)
3 boundary up $u_n$ tetrahedra share 6 (blue) triangles with 3 down tetrahedra $d_n$.}
\end{center} 
\end{figure}

We use also the notation $t_{nm4}\equiv (n,m,4)$,  $t_{nm5}\equiv (n,m,5)$, and the notation $t^{\rm u,d}_1 \equiv (2,3,4,5)_{u,d}$, $t^{\rm u,d}_2 \equiv (3,1,4,5)_{u,d}$, $t^{\rm u,d}_3 \equiv (1,2,4,5)_{u,d}$ (cyclically in 1,2,3).  The tetrahedra in the triangulation $\Delta_6$ and their relations are represented in Fig.\ref{diag2}. 

For this term, we have 
\begin{equation}
W[s_6]=
\begin{tiny}  
\sum_{j} \dim(j)\prod_{n<m}\dim(j^{\rm u}_{nm})\dim(j^{\rm d}_{nm})\dim(j^{\rm s}_{n4})\dim(j^{\rm s}_{n5})\hspace{1em}\mathcal{B}(j,j^{u}_{nm},j^{\rm s}_{nv})\hspace{1em}Ê\mathcal{B}(j,j^{d}_{nm},j^{\rm s}_{n}). 
\end{tiny}
\end{equation}
Notice the sum over the spin $j$ of the internal face. This sum is finite, because it is controlled by the Clebsch-Gordan relations between spins at the edges. 

The vacuum boundary state $\Psi_{\mathbf q}[s_6]$ for  spin network $s_6$ will be a function 
$\Psi_{\mathbf q}[s_6]$ peaked on background values $( j^{\rm u,d(6)}_{nm},
 j^{\rm s(6)}_{nv})$ which represent 
a given background geometry $\mathbf q_6$.  Notice that we cannot imbed 
two flat non-degenerate 4-simplices glued along \emph{two} faces into $R^4$ (for the same reason for which two non-degenerate triangles in $R^2$ cannot be glued along
\emph{two} sides).  Thus, we cannot fix the geometry $\mathbf q_6$ as we did in the  previous case.  Instead, let us proceed as follows. 

Let $v_{\rm u}$ and $v_{\rm d}$ be two regular 4-simplices, of side $L$. Identify two tetrahedra ($p_{\rm 4}$ and  $p_{\rm 5}$) of these two 4-simplices.  This defines a conical space that is flat except on the triangle $\tau$ that separates $p_{\rm 4}$ and  $p_{\rm 5}$, where the deficit angle is 2$\pi$ minus twice the dihedral angle of a regular 4-simplex.   This space has a fixed boundary geometry, which we take as the definition of $\mathbf q_6$. 

We take the boundary state to be a Gaussian peaked on  $\mathbf q_6$:
\begin{equation}
\Psi_{\mathbf q}[s_6] = 
C_6 \ e^{
- \alpha_{ll'}{(j_l- j_l^{\scriptscriptstyle (6)})(j_{l'}- j_{l'}^{\scriptscriptstyle (6)})}
+i \Phi_l^{\scriptscriptstyle (6)}j_{l}}. 
\label{vuoto2}
\end{equation} 

Following the same steps as above, we found now that the Regge action is now also
a function of the summation variable $j_{45}$, which represents the area of the internal triangle $\tau$. 
\begin{eqnarray}
\label{regges_6}
S_{Regge}({\scriptstyle j_{nm}^{\rm u},j_{nm}^{\rm d}, j^{\rm l}_{n}, j^{\rm r}_{n},j}) &=& \tilde \phi^{(6)}_{n m} j_{n m}^{\rm u} +
 \tilde \phi^{(6)}_n j^{\rm s}_{n,v} + 
\tilde \phi^{(6)}_{n m} j_{n m}^{\rm d} +
 \tilde \phi^{(6)}_n j^{\rm s}_{n,v} + \\ \nonumber 
&+& \frac{1}{2} G_{ll'} \, \delta j_l \delta j_{l'} 
+ G_{(45)l'} \, \delta j_{45} \delta j_{l'}+ \\ \nonumber 
&+& (\tilde \phi^{u(6)}_{45} +\tilde \phi^{d(6)}_{45})j_{45} +\frac{1}{2} G_{(45)(45)} \, \delta j_{45} \delta j_{45}, 
\end{eqnarray}
where $\tilde \phi^{(6)}_n$ and $\tilde \phi^{(6) }_{n m}$ are the dihedral angles of flat 4-simplices 
with the given boundary geometry and are supposed to be function of the reference background value $j^{(6)}_{45}$,
 like the ``discrete derivative" and the fluctuation $\delta j_{45}= j_{45}-j^{(6)}_{45}$.
Since we have an internal loop we sum over $\delta j_{45}$
\begin{eqnarray}
{\mathbf G_{s_6}}({L'},T)=\frac{4 \Lambda^2 {\cal N}_6 }{j_{TL}^2} \, 
 \sum_{\delta j_{l}}\sum_{\delta j_{45}} 
\delta j^{\rm u}_{l} \,  \delta j^{\rm d}_{l'}  \, 
P_{\tau}^2 \, e^{-i\left(S_{Regge} 
+ k_{\tau} \frac{\pi}{2}\right)} e^{
-\tilde\alpha_{ll'}(\delta j_l)(\delta j_{l'})
+ i  \Phi^{(6)} j_l}
\end{eqnarray} 
where ${\cal N}_6$ is fixed by the normalization. The first order term of the expansion of the Regge action cancels the phases in the boundary state but gives 
an extra phase $i\Big(\tilde \phi^{(6)u}_{45} + \tilde \phi^{(6)d}_{45}\Big)j_{45}$ and two discrete derivative 
terms $G_{(45)l}$ and $G_{(45)(45)}$ related to the internal loop, leaving
\begin{eqnarray}
{\mathbf G_{s_6}}({L'},T)&=&N_6
 \sum_{\delta j_{l}}\sum_{\delta j_{45}} 
\delta j^{\rm u}_{l} \,  \delta j^{\rm d}_{l'} \,
exp\Big(-\tilde\alpha_{ll'}\delta j_l \, \delta j_{l'}
- \frac{i}{2} G_{ll'} \, \delta j_l \, \delta j_{l'}+ \\ \nonumber
 &+& i \, (\tilde \phi^{(6)u}_{45} + \tilde \phi^{(6)d}_{45})j_{45} +
G_{(45)l'} \, \delta j_{45} \delta j_{l'}+
\frac{1}{2} G_{(45)(45)} \, \delta j_{45} \delta j_{45}\Big)
\end{eqnarray} 
The dihedral angle $\tilde \phi^{(6)u}_{45}$ are the angles between the normals of the two internal tetrahedra 
($p_1$ and $p_2$) in the two 4-simplices.  Their sum is the deficit angle at the triangle
${\cal T}_2$. As mentioned, this deficit angle cannot be zero (or a multiple of 2$\pi$), because there is no imbedding of two nondegenerate 4-simplices glued by \emph{two} tetrahedra into $R^4$. Therefore 
\begin{equation}
\tilde \phi^{(6)u}_{45} + \tilde \phi^{(6)d}_{45} \ne 0.
\end{equation}
But it follows from this that the sum over $\delta_{45}$ is a sum of a rapidly oscillating function, and is therefore suppressed.  Thus, this term is strongly suppressed in the large $j$ limit.  Notice that the denominator might also be suppressed at this order in $\lambda$, but is not going to be suppressed at all orders in $\lambda$; therefore the suppression is effective.

\subsection{Three internal propagators:  $2\rightarrow 3\rightarrow2$ Pachner's move}

\begin{figure}[b]
\begin{center}
\includegraphics[height=5cm]{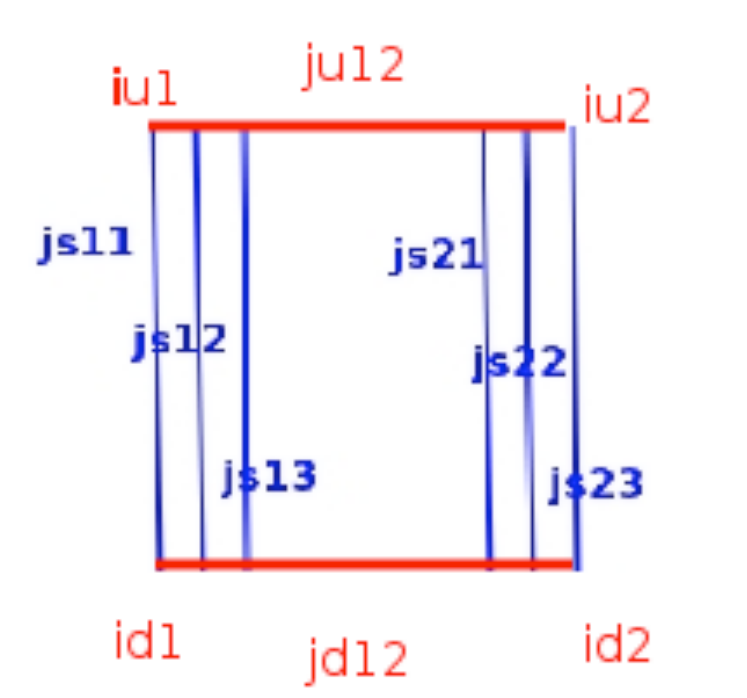}
\caption{The boundary spin network $s_4$.}
\label{spinet4}
\end{center}
\end{figure}

\noindent Consider the Feynman graph
 \begin{center}
  \includegraphics[height=2cm]{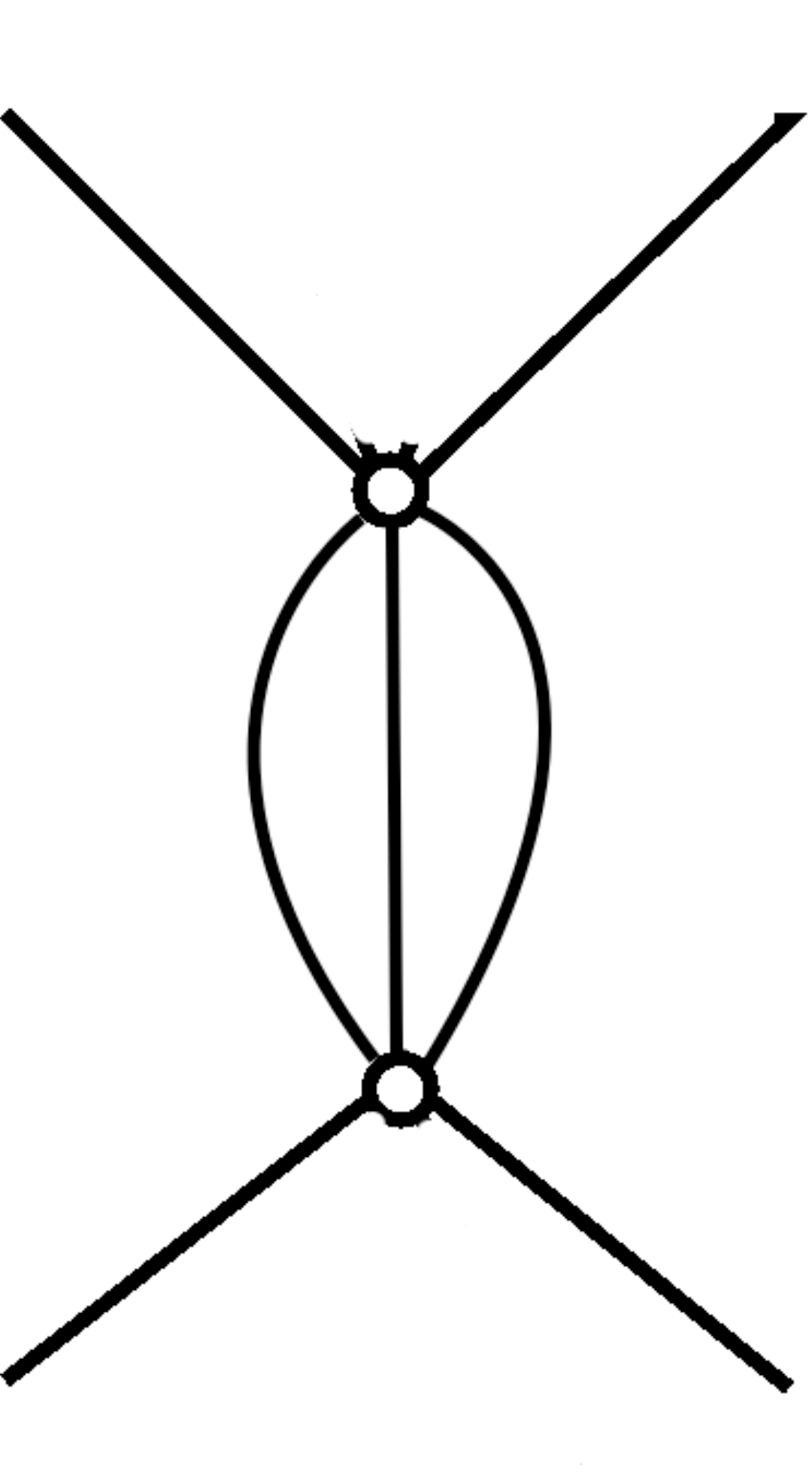} \hspace{1em} .
  \end{center}
  The boundary graph $\Gamma_4$ is illustrated in Fig.\ref{spinet4}: two theta spin networks connected by two links.  

\begin{figure}
\begin{center}
\includegraphics[height=8 cm]{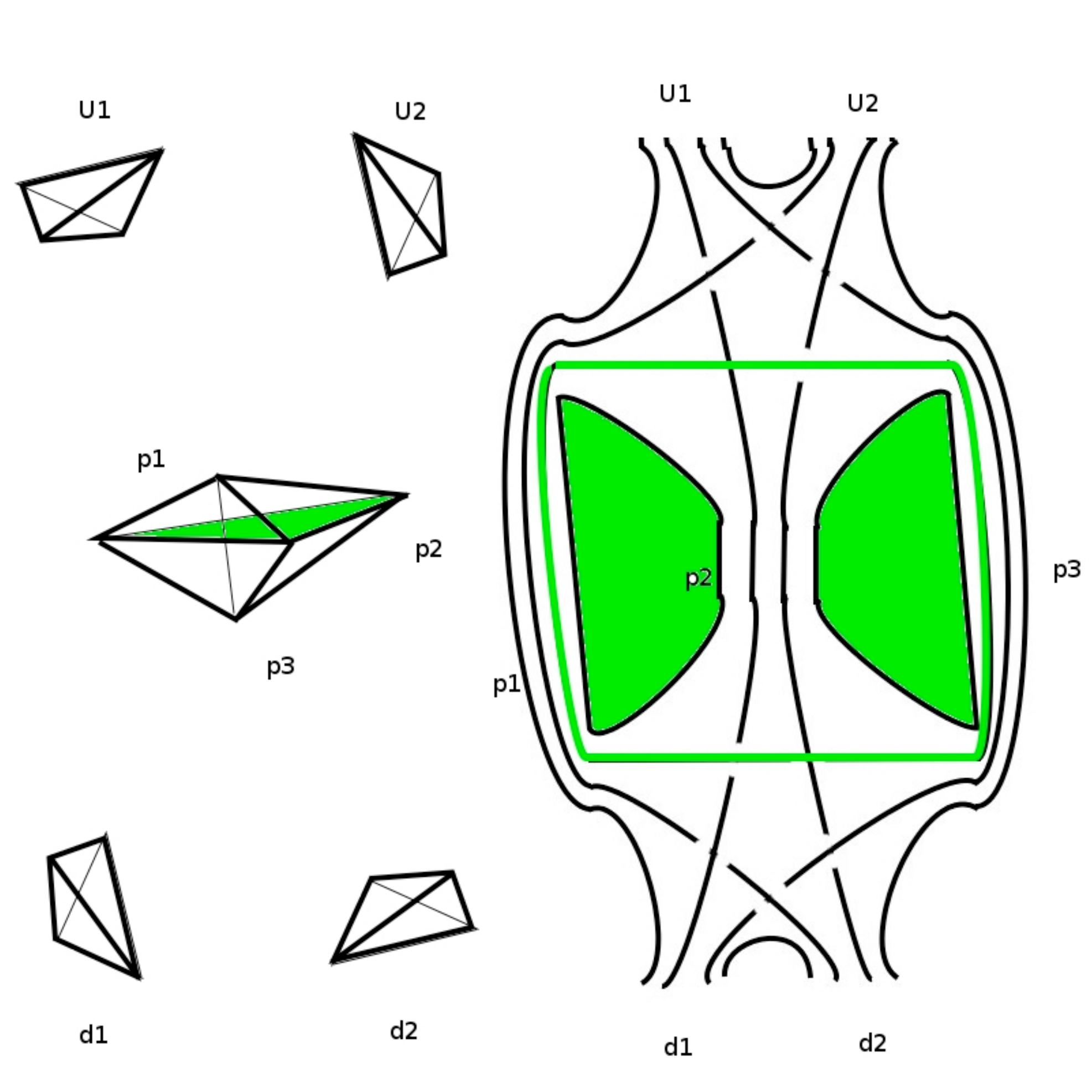}
\raisebox{4cm}{
\begin{tabular}{ccccc}
& & & & \\ \hline
\multicolumn{1}{|c|}{Tetrahedra} & \multicolumn{4}{|c|}{Triangles} \\
\hline
& & & &  \\
\hline
\multicolumn{1}{|c|}{$u_1$}  & \multicolumn{1}{c}{\textcolor{red}
{$t^{\rm u}$}}  & \multicolumn{3}{c|}{\textcolor{blue}
{$t_{234}$} \comments{\textcolor{blue}}
{$t_{235}$} \comments{\textcolor{blue}}
{$t_{245}$}} \\
 \hline
\multicolumn{1}{|c|}{$u_2$} & 
\multicolumn{1}{c}{\textcolor{red}
{$t^{\rm u}$}}  &
\multicolumn{3}{c|}{\textcolor{blue}
{$t_{134}$} \comments{\textcolor{blue}}
{$t_{135}$} \comments{\textcolor{blue}}
{$t_{145}$}} \\ \hline
& & & &  \\
& & & &  \\
\hline

\multicolumn{1}{|c|}{$p_1$} & \comments{\textcolor{blue}}
{$t_{134}$} & \comments{\textcolor{blue}}
{$t_{234}$} &  \multicolumn{2}{c|}{\textcolor{green}
{$t_{123}$} \comments{\textcolor{green}}
{$t_{124}$}} \\ \hline
\multicolumn{1}{|c|}{$p_2$} & \comments{\textcolor{blue}}
{$t_{135}$} & \comments{\textcolor{blue}}
{$t_{235}$} & \multicolumn{2}{c|}{\textcolor{green}
{$t_{123}$} \comments{\textcolor{green}}
{$t_{125}$}} \\ \hline
\multicolumn{1}{|c|}{$p_3$} & \comments{\textcolor{blue}}
{$t_{145}$} & \comments{\textcolor{blue}}
{$t_{245}$} & \multicolumn{2}{c|}{\textcolor{green}
{$t_{124}$} \comments{\textcolor{green}}
{$t_{125}$}} \\ \hline
& & & &  \\
& & & &  \\
\hline
\multicolumn{1}{|c|}{$d_1$}  & \multicolumn{1}{c}{\textcolor{red}
{$t^{\rm d}$}}  & \multicolumn{3}{c|}{\textcolor{blue}
{$t_{234}$} \comments{\textcolor{blue}}
{$t_{235}$} \comments{\textcolor{blue}}
{$t_{245}$}} \\
 \hline
\multicolumn{1}{|c|}{$d_2$} & 
\multicolumn{1}{c}{\textcolor{red}
{$t^{\rm d}$}}  &
\multicolumn{3}{c|}{\textcolor{blue}
{$t_{134}$} \comments{\textcolor{blue}}
{$t_{135}$} \comments{\textcolor{blue}}
{$t_{145}$}} \\ \hline
& & & &  \\
\multicolumn{5}{l}{\small{The labels $u$ and $d$ refer to $345$.}} \\
\end{tabular}}
\caption{{\it Tetrahedral decomposition}: two 4-simplices share 
three tetrahedra $p_1$, $p_2$ and $p_3$.}
\label{diagramma3}
\end{center}
\end{figure}

Denote the nodes of the first link as $u_1$ and $u_2$ and the nodes of the second one as $d_1$ and $d_2$.   
The generalized triangulation $\Delta_4$ that gives the maximal contribution is obtained by gluing two 4-simplices via \emph{three} tetrahedra. See Figure \ref{diagramma3}. 

\begin{figure}[h]
  \begin{center}
  \includegraphics[height=4cm]{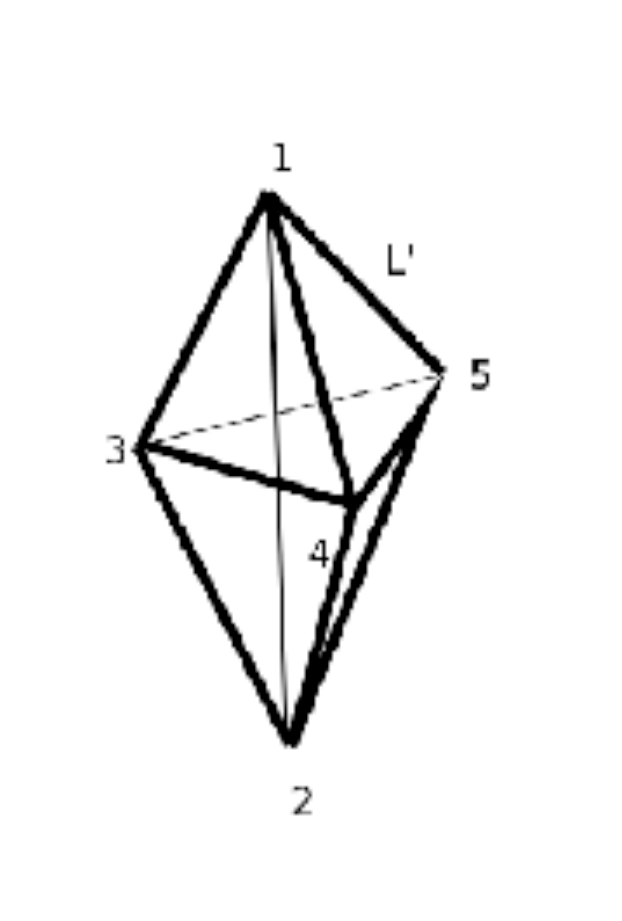}
\vspace{-1cm}
  \end{center}
  \caption{\label{tetrabordo3} Labelling of the vertices of $\Delta_4$.}
  \end{figure}

The triangulation $\Delta_4$ is formed by 5 points, which we label as $1,2,3,4,5$; by the 10 edges ${(n,m),(n,1),(n,2), (1,2)}$ (here $n=3,4,5$); the 11 faces $(n,m,1),(n,m,2),(n,1,2), (3,4,5)_u, (3,4,5)_d$  (notice that there are two distinct  \emph{triangles} connecting the points 3, 4  and 5); the 7 tetrahedra  $(n,m,1,2), (3,4,5,1)_u,(3,4,5,2)_u,(3,4,5,1)_d,(3,4,5,2)_d$; and the two 4-simplices  $(1,2,3,4,5)_u,(1,2,3,4,5)_d$. See Figure \ref{tetrabordo3}. 
 
We use the notation $p_3=(4,5,1,2), p_4=(5,3,1,2), p_5=(3,4,1,2)$ and we call $\mathcal{T}_3=p_3\cup p_4\cup p_5$ the union of the three central tetrahedra. 
The triangulation $\Delta_4$ can be seen as the world-history of the triangle $(3,4,5)_d$ evolving in to   $\mathcal{T}_3$ and then recollapsing into the triangle  $(3,4,5)_u$.   The initial and final triangles share their perimeters, and in particular their vertices.
 
The representations associated to the 8 links $l^{\rm u}_{12},l^{\rm d}_{12}, l^s_{nv}$ of $\Gamma_4$ are $j^{\rm u}_{12}$, 
$j^{\rm d}_{12}$, $j^s_{nv}$ while $i_n^{\rm u}, i_n^{\rm d}$ are the
four intertwiners associated to the four nodes $u_n$ and $d_n$ ($n=1,2$).  
The set $s_4=(\Gamma_4,  j^{\rm u}_{12},j^{\rm d}_{12},j^{\rm s}_{nv}, i_n^{\rm u}, i_n^{\rm d})$ 
is the boundary spin network we consider in this section.  The boundary function $f_{s4}(\phi)$ is of order four
\begin{equation} 
f_{s_4}(\phi) = \sum_{\alpha_{nm}\beta_{nm}}\prod_{n=1,2} 
\phi^{\alpha_{nm}  i^{\rm u}_n}
_{j^{\rm u}_{nm}}\ 
\phi^{\beta_{nm}  i^{\rm d}_n}
_{j^{\rm d}_{nm}}
\end{equation}
and the expansion gives two vertices and seven propagators: 
\begin{eqnarray}
W[s_4] &=& 
\begin{tiny}  
\Big(\prod_{n=1,2}\mathcal{P}_{\alpha_{nm}\alpha_{n}i^{\rm u}_n}^{j^{\rm u}_{nm}\,j_n}\,{}_{\alpha'_{nm}i'_n}^{j'_{nm}}\Big)
\mathcal{V}_{\alpha'_{nm}\gamma_m\eta_mi'_ni''i^{IV}i^{VI}}^{j'_{nm}j''_mj^{IV}_mj^{VI}_m}
\mathcal{P}_{\gamma_mi''}^{j''_m}\,{}_{\delta_mi'''}^{j'''_m}
\mathcal{P}_{\gamma_mi^{IV}}^{j^{IV}_m}\,{}_{\delta_mi^{V}}^{j^{V}_m}
\end{tiny}
 \\ \nonumber
&&
\begin{tiny}  
\times \ \ \ \ \mathcal{P}_{\eta_mi^{VI}}^{j^{VI}_m}\,{}_{\theta_mi^{VII}}^{j^{VII}_m}  
\mathcal{V}_{i^{VII}\theta_mi^{V}\delta_{m}i'''\beta'_{nm}i'_n}^{j^{VII}_mj^{V}_mj'''_{m}j''_{nm}}  \Big(\prod_{n=1,2}\mathcal{P}_{\beta_{nm}\beta_{n}i^{\rm d}_n}^{j^{\rm d}_{nm}\,j_n}\,{}_{\beta'_{nm}i'_n}^{j''_{nm}}\Big) 
\end{tiny}
\\ \nonumber
&=&
\begin{tiny}  
 \prod_{n=1,2}\dim(i^{\rm u}_{n})\dim(i^{\rm d}_{n})\prod_{v=1,3}\dim(j_{n,v})
\dim(j^{\rm u}_{12})\dim(j^{\rm d}_{12})\sum_{j_{kl}} 
\Big(\prod_{\scriptscriptstyle k=3,4 \, k<l}\dim(j_{kl})\Big)
\mathcal{B}(j_{nm}^{(4)})\mathcal{B}(j_{nm}^{(4)}) 
\end{tiny}
\end{eqnarray} 
where the $I$ label of $j_I$ refer to the three internal faces shared by the tetrahedra, 
along which they are glued together (loops in the GFT diagram).

We now chose a boundary geometry with nondegenerate areas and dihedral angles,
defining nondegenerate 4-simplices. Let  $j_l=(j_{12}^{\rm u},j_{12}^{\rm d}, j^{\rm s}_{nv})$ be the background values on which the boundary function $\Psi_{\mathbf q}[s_4]$ is peaked: 
\begin{equation}
\Psi_{\mathbf q}[s_4] = 
C_4 \ e^{
- (\alpha_4)_{ll'}{(j_l- j_l^{\scriptscriptstyle (4)})(j_{l'}- j_{l'}^{\scriptscriptstyle (4)})}
+i \Phi_l^{\scriptscriptstyle (4)}j_{l}}
\label{vuoto3}
\end{equation} 
and $( j^{\rm u,d(4)}_{nm}, j^{\rm s(4)}_{n})$ the areas of the internal faces determined by the boundary geometry.  The Regge action, as before, will be written as an expansion over the boundary $j_l$ but also over the internal background reference $j^{(4)}_{34}$,$j^{(4)}_{35}$ and $j^{(4)}_{45}$:
\begin{eqnarray}
\hspace{-1cm}
S_{Regge}({\scriptstyle j_{nm}^{\rm u}, j_{nm}^{\rm d}, j^{\rm s}_{n,v}, j_{IJ}}) &=& \tilde \phi^{(4)}_{n m} j_{n m}^{\rm u} +
 \tilde \phi^{(4)}_n j^{\rm s}_{n,v} + 
\tilde \phi^{(4)}_{n m} j_{n m}^{\rm d} +
 \tilde \phi^{(4)}_n j^{\rm s}_{n,v} +  \\ \nonumber
&+& \frac{1}{2} G_{ll'} \, \delta j_l \delta j_{l'}+ 
+ i\Big(\tilde \phi^{u(4)}_{IJ}+\tilde \phi^{d(4)}_{IJ}\Big)j_{IJ} + \\ \nonumber
&+& G_{(IJ)l'} \, \delta j_{IJ} \delta j_{l'}+ \frac{1}{2} G_{(IJ),(KL)} \, \delta j_{IJ} \delta j_{KL} 
\end{eqnarray}
where $I,J,K,L=3,4,5$, $I<J$, $K<L$, $\tilde \phi^{(4)}_{n,v}$ and $\tilde \phi^{(4) }_{n m}$ are the dihedral angles. 

The phases of the external angles cancel the phases in the boundary state. The phases in the internal angles are 
\begin{equation}
\Big(\tilde \phi^{(4)u}_{34} + \tilde \phi^{(4)d}_{34}\Big)j^{4}_{34} + \Big(\tilde \phi^{(4)u}_{35} + 
\tilde \phi^{(4)d}_{35}\Big)j^{4}_{35} + \Big(\tilde \phi^{(4)u}_{45} + \tilde \phi^{(4)d}_{45}\Big)j^{4}_{45}=0
\end{equation}
The sum over the three independent variables $j_{34}, j_{45}, j_{53}$ suppresses again the amplitude, because the deficit angles in the parenthesis cannot vanish, for the same reason as in the previous section. 

However, this result raise a problem.  We expect contributions in a Feynman sum to be suppressed in the semiclassical approximation if there is no classical trajectory that give a saddle point in the sum. Here the classical trajectories should reproduce the Einstein equations, and these demand the Ricci tensor to vanish, and not the Riemann tensor. But the vanishing of all deficit angles above correspond to to flatness, namely to the vanishing of the Riemann tensor.  Why don't (non-flat) Ricci-flat configurations contribute in the semiclassical limit? 

The origin of the problem can be traced to the oversimplification of the dynamics which characterizes the old Barrett-Crane vertex.  In this model, the spins are the sole dynamical variables.  In the semiclassical limit, they correspond to areas of triangles of Regge-like triangulations.  The vertex approximates correctly the Regge action, but this is not sufficient to reproduce the Regge-calculus dynamics, because the variables are areas instead of lengths. On this, see \cite{Dittrich:2007wm}.
  Let's see how this problem reflects here.  

We are considering a (generalized) triangulation obtained by gluing two 4-simplices by three tetrahedra.  The segments of this triangulation are 10, because all segments are shared by the two 4-simplices.   But the areas are 11, because only 9 triangles are shared, and each 4-simplex has a triangle which is not shared with the other 4-simplex. Therefore there should be one relation among the areas to be satisfied if the areas have to define a geometrical (Regge-like) generalized triangulation.  Of these 11 areas, 8 are boundary areas, and three are internal. If we fix the 8 external areas, there should be a relation between the three internal areas.  Suppose we linearize this relation
\begin{equation}
a \delta j_{34} + b \delta j_{45} + c \delta j_{53} = 0
\end{equation}
and impose this in the integral of $\delta j_{nm}$. Then the integral is not anymore suppressed, provided that the deficit angle angles satisfy a relations like 
\begin{equation}
\Big(\tilde \phi^{(4)u}_{34} + \tilde \phi^{(4)d}_{34}\Big) =a \phi, \hspace{2em}
\Big(\tilde \phi^{(4)u}_{45} + \tilde \phi^{(4)d}_{45}\Big) =b \phi, \hspace{2em}
\Big(\tilde \phi^{(4)u}_{53} + \tilde \phi^{(4)d}_{53}\Big) =c \phi . 
\end{equation}
That is, the amplitude may fail to be suppressed even if the triangulation is not flat.  It is reasonable to expect that the above condition reflects Ricci flatness.

In the model we are considering, a relation between the fluctuations of the spins does not seem to be implemented; and this is perhaps one additional sign of the problems of the old Barrett-Crane model.   Do the new models correct this problem?

\subsection{Four internal propagators:  $1 \rightarrow 4 \rightarrow 1$ Pachner's move}

Finally, consider the Feynman graph 
 \begin{center}
  \includegraphics[height=4cm]{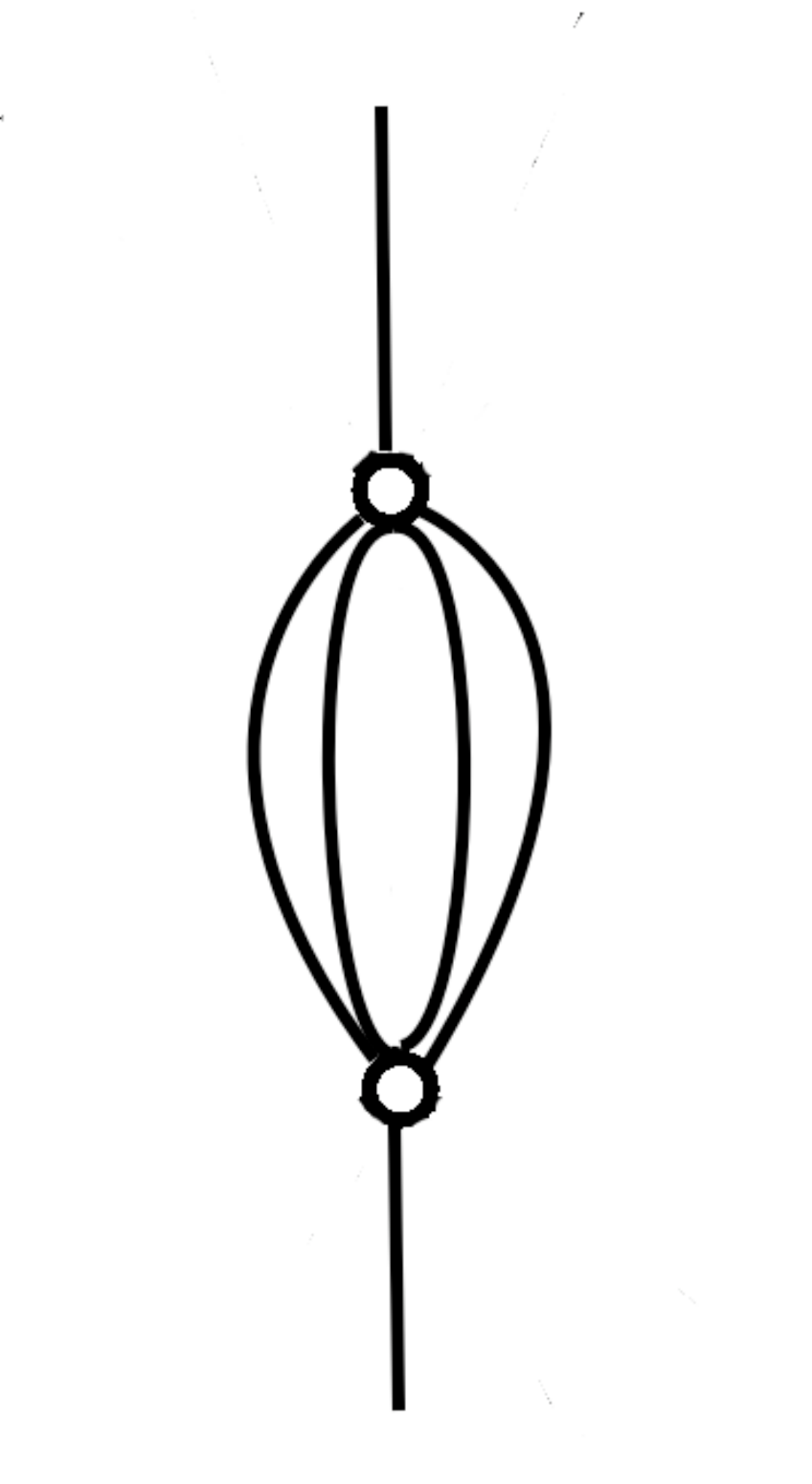}
  \end{center}
which looks like a self-energy correction for the GFT propagator.  The potential divergences of this graph have been analyzed in \cite{perini} in the simple case of vanishing boundary areas.

\begin{figure}[jh]
\begin{center}
\includegraphics[height=5cm]{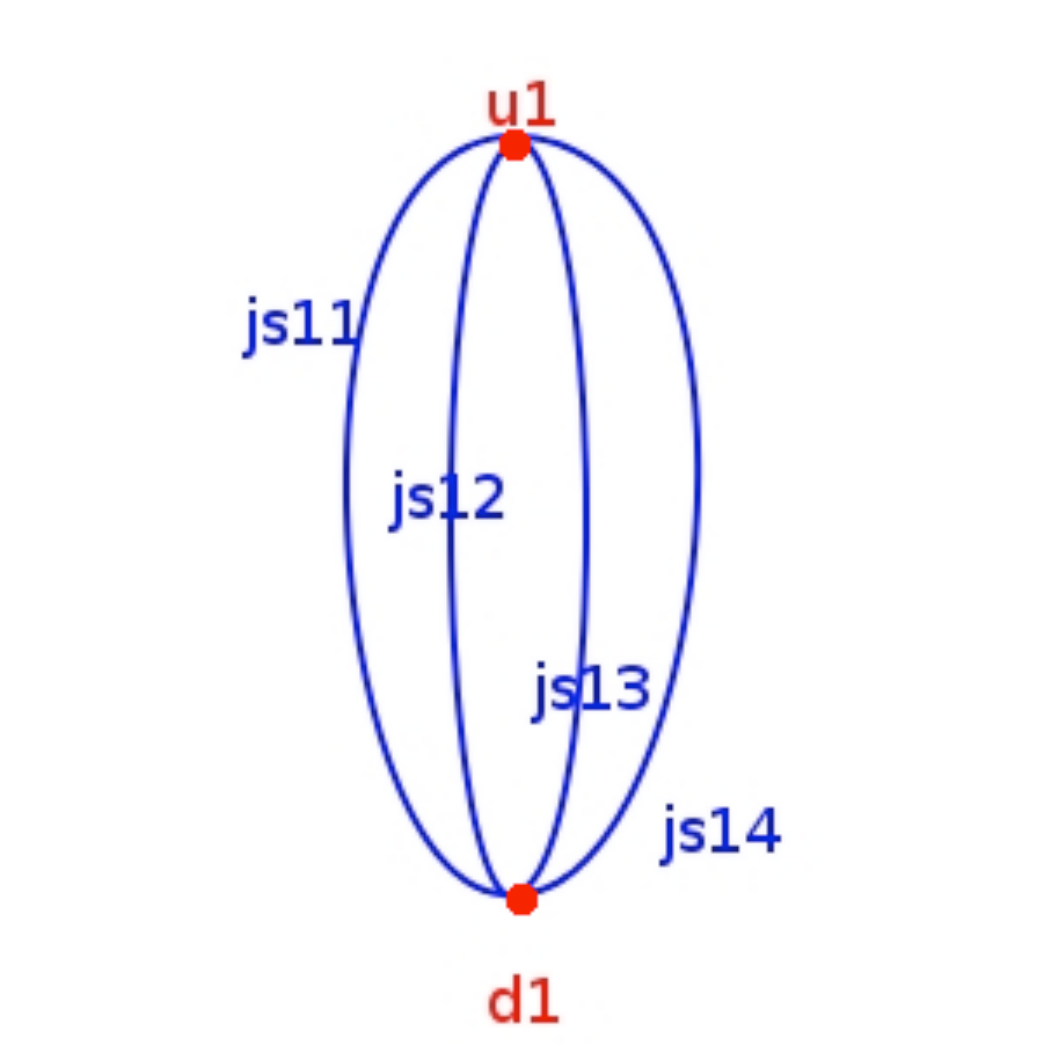}
\caption{The boundary spin network $s_2$.} 
\label{spinet44}
\end{center}
\end{figure}

The boundary graph $\Gamma_2$ is dual to the spinnetwork of Fig.\ref{spinet44}: a tetrahedral spin network with two nodes $u_1$ and $d_1$. The links of $\Gamma_2$ are the four side links $l^{\rm s}_n$, which connect 
$u_1$ with $d_1$.

The corresponding maximal four--dimensional triangulation $\Delta_2$ is made by two 4-simplices glued by \emph{four} tetrahedra.

\begin{figure}[h]
  \includegraphics[height=3cm]{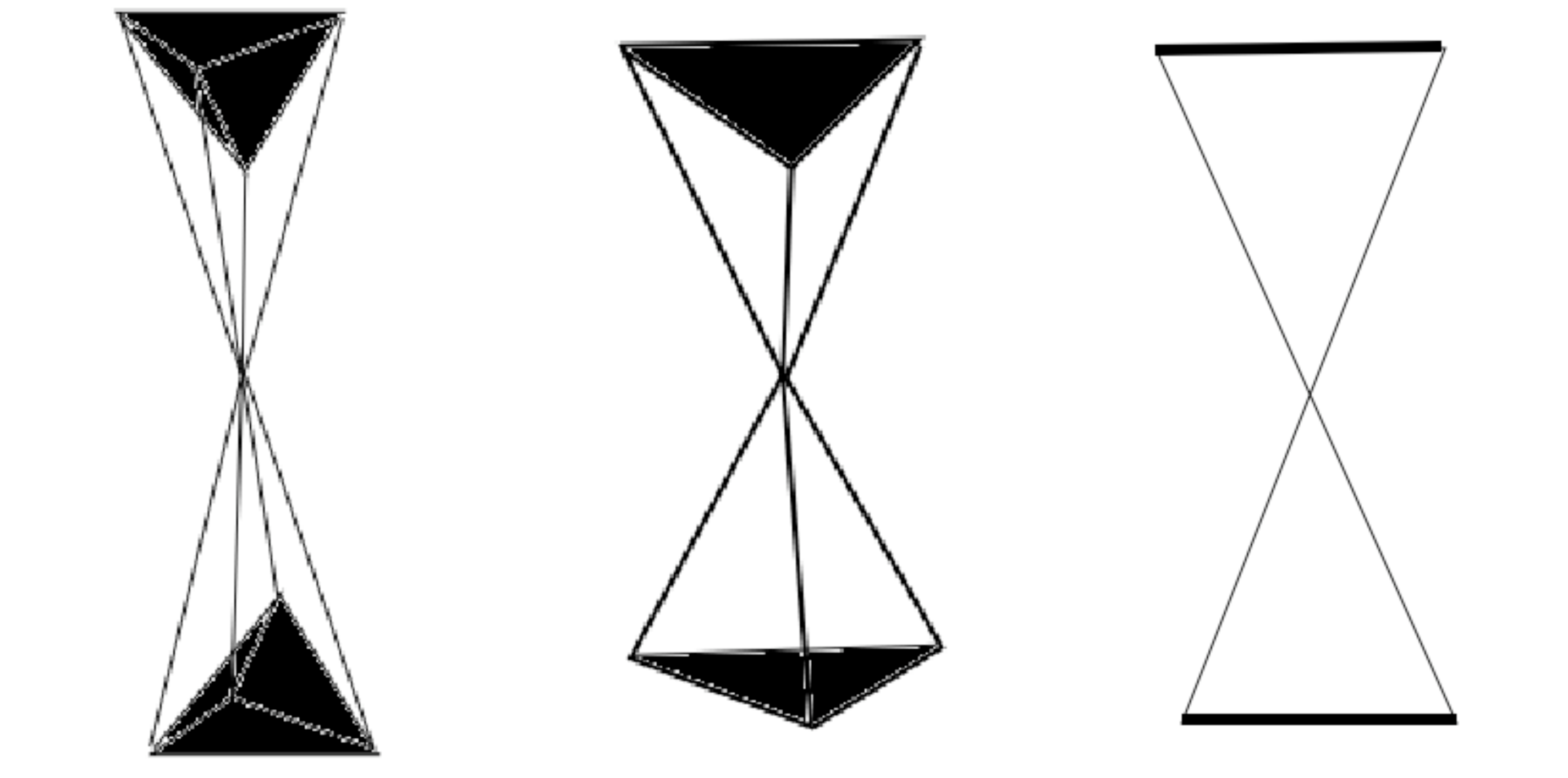}
  \caption{\label{annichila} The spacetime triangulation $\Delta_2$ and the lower dimensional analogies. Upper and lower lateral side must be identified.}
  \end{figure}

The triangulation $\Delta_2$ is formed by 5 points, which we label as $1,2,3,4,5$; by the 10 edges ${(n,m),(n,5)}$ (here $n=1,2,3,4$); the 10 faces $(n,m,n),(n,m,5)$; the 6 tetrahedra  $(n,m,p,5), (1,2,3,4)_u,  (1,2,3,4)_d$  (notice that there are two distinct  \emph{tetrahedra} connecting the points 1, 2, 3  and 4); and the two 4-simplices  $(1,2,3,4,5)_u,(1,2,3,4,5)_d$. See Figure \ref{annichila}. 
 
This can be seen as the world-history of the tetrahedron $d \equiv (1,2,3,4)_d$ evolving into a set of four tetrahedra, having the same 3d boundary of original one, and then evolving back to a single tetrahedron $u \equiv  (1,2,3,4)_u $.  

\begin{figure}[b]
  \begin{center}
  {\includegraphics[height=3cm]{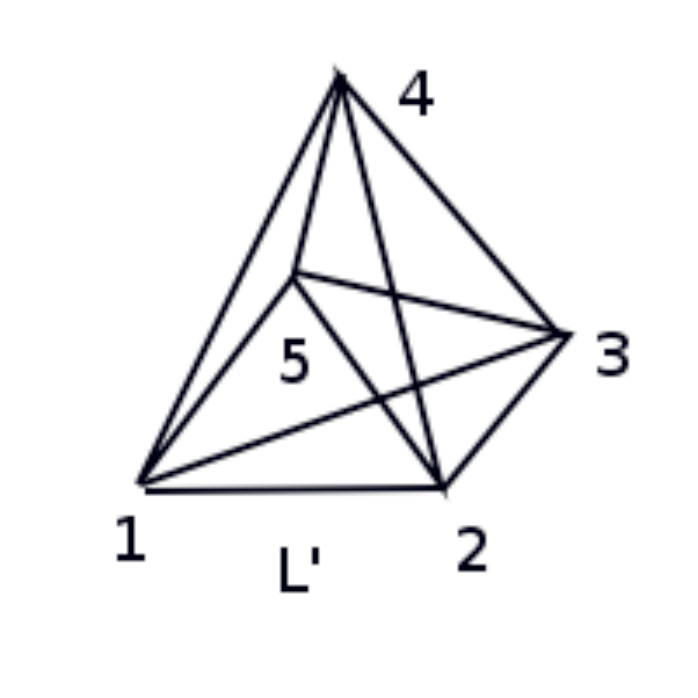}}
  \end{center}
  \vspace{-1cm}
  \caption{\label{tetrabordo4} The labeling of the vertices of the triangulation $\Delta_2$.}
  \end{figure}

\begin{figure}
\begin{center}
\includegraphics[height=8 cm]{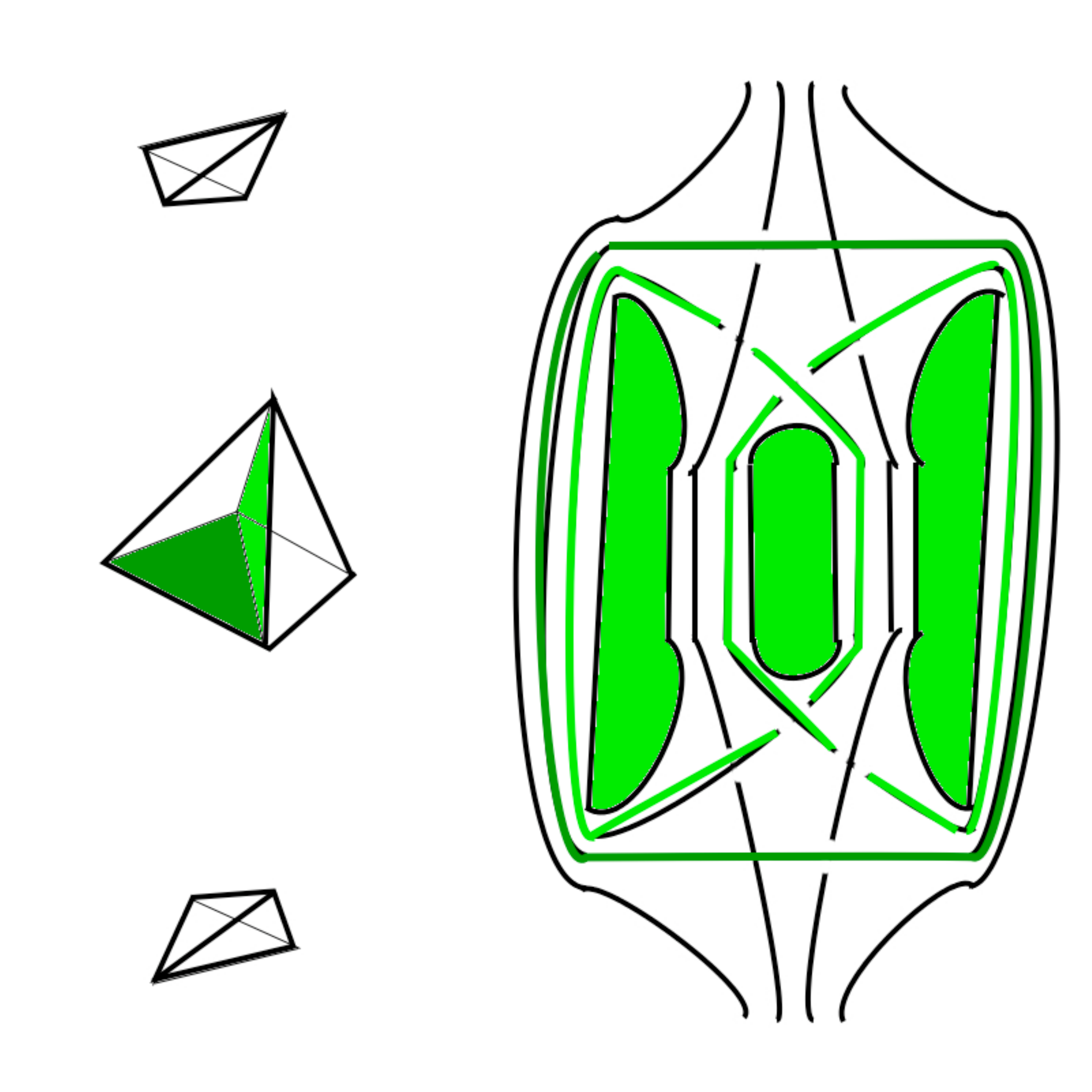}
\hspace{3em}
\raisebox{4cm}{\begin{tabular}{ccccc}
& & & & \\ \hline
\multicolumn{1}{|c|}{Tetrahedra} & \multicolumn{4}{|c|}{Triangles} \\
\hline
& & & &  \\
\hline
\multicolumn{1}{|c|}{$u_1$}  & 
\multicolumn{4}{c|}{\textcolor{blue}
{$t_{234}$}   \comments{\textcolor{blue}}
{$t_{235}$}  \comments{\textcolor{blue}}
{$t_{245}$}  \comments{\textcolor{blue}}
{$t_{345}$}} \\
 \hline

& & & &  \\
& & & &  \\
\hline
\multicolumn{1}{|c|}{$p_1$} & \comments{\textcolor{blue}}
{$t_{345}$} & \multicolumn{3}{c|}{\textcolor{green}
{$t_{134}$} \comments{\textcolor{green}}
{$t_{135}$}  \comments{\textcolor{green}}
{$t_{145}$}} \\ \hline
\multicolumn{1}{|c|}{$p_2$} & \comments{\textcolor{blue}}
{$t_{245}$} & \multicolumn{3}{c|}{\textcolor{green}
{$t_{124}$}  \comments{\textcolor{green}}
{$t_{125}$} \comments{\textcolor{green}}
{$t_{145}$}} \\ \hline
\multicolumn{1}{|c|}{$p_3$} & \comments{\textcolor{blue}}
{$t_{235}$} & \multicolumn{3}{c|}{\textcolor{green}
{$t_{123}$}  \comments{\textcolor{green}}
{$t_{125}$} \comments{\textcolor{green}}
{$t_{135}$}} \\ \hline
\multicolumn{1}{|c|}{$p_4$} & \comments{\textcolor{blue}}
{$t_{234}$} & \multicolumn{3}{c|}{\textcolor{green}
{$t_{123}$}  \comments{\textcolor{green}}
{$t_{124}$} \comments{\textcolor{green}}
{$t_{134}$}} \\ \hline
& & & &  \\
& & & &  \\
\hline
\multicolumn{1}{|c|}{$d_1$}  & 
\multicolumn{4}{c|}{\textcolor{blue}
{$t_{234}$}   \comments{\textcolor{blue}}
{$t_{235}$}  \comments{\textcolor{blue}}
{$t_{245}$}  \comments{\textcolor{blue}}
{$t_{345}$}} \\
 \hline

& & & &  \\
\end{tabular}
}
\label{diagramma4}
\caption{The tetrahedral decomposition of the $1\to 4 \to 1$ diagram.}
\end{center}
\end{figure}

 The \emph{boundary} of $\Delta_2$ is made by two tetrahedra (Fig.\ref{diagramma4}): ``up" tetrahedron $u_1$ and ``down" $d_1$. They share all their four faces, that is, the triangles $t_{234}$, $t_{235}$, $t_{245}$, $t_{345}$. See Figure \ref{tetrabordo4}. 

Denote $j^{\rm s}_{v}$ ($v=1,4$) the spins associated to the 4 links 
$l^s_{n}$ of $\Gamma_2$, and $i_1^{\rm u}, i_1^{\rm d}$ the two intertwiners associated
 to the two nodes $u_1$ and $d_1$.  
The set $s_2=(\Gamma_2, j^{\rm s}_{v}, i_1^{\rm u}, i_1^{\rm d})$ with $v=1,2,3,4$ is the boundary spin network.  The boundary function $f_{s_2}(\phi)$ is of order two 
\begin{equation} 
f_{s_2}(\phi) = \sum_{\alpha_{nm}\beta_{nm}} 
\phi^{\alpha_{nm}  i^{\rm u}_1}
_{j^{\rm u}_{nm}}\ 
\phi^{\beta_{nm}  i^{\rm d}_1}
_{j^{\rm d}_{nm}}
\end{equation}
 The Wick expansion of the highest dimensional contribution gives two vertices and six propagators
\begin{eqnarray}
\hspace{-5cm}
W[s_2] &=& \begin{tiny}  
\mathcal{V}_{\alpha'_{nm}\gamma_m\eta_m i'_n i^{II}i^{IV}i^{VI}
 i^{VIII}}^{j'_{nm}j^{II}_m j^{IV}_m j^{VI}_m j^{VIII}_m}
\mathcal{P}_{\gamma_mi^{II}}^{j^{II}_m},\,{}_{\delta_mi^{III}}^{j^{III}_m} 
\end{tiny}
\\ \nonumber
&& \ \  \times \ \begin{tiny}  
 \mathcal{P}_{\gamma_mi^{IV}}^{j^{IV}_m},\,{}_{\delta_mi^{V}}^{j^{V}_m}
\mathcal{P}_{\eta_mi^{VI}}^{j^{VI}_m},\,{}_{\theta_mi^{VII}}^{j^{VII}_m}
\mathcal{P}_{\eta_mi^{VIII}}^{j^{VIII}_m},\,{}_{\theta_mi^{IX}}^{j^{IX}_m}
\mathcal{V}_{i^{IX}i^{VII}\theta_mi^{V}\delta_{m}
i^{III}\beta'_{nm}i'_n}^{j^{VII}_mj^{V}_mj^{III}_{m}j^{II}_{nm}}
\end{tiny}
\\ \nonumber 
&=& \begin{tiny}  
 (\dim i^{\rm u}_{1}\dim i^{\rm d}_{1})\Big(\prod_{v=1,4}\dim(j_{1v})\Big)
\sum_{j_{KL}}\Big(\prod_{\scriptscriptstyle K=2,3,4,5;K<L}\dim(j_{KL})
\mathcal{B}(j_{nm})\mathcal{B}(j_{nm})\Big) \end{tiny}
\end{eqnarray} 

\vspace{0.5 cm}

Let us choose again the boundary geometry that defines the boundary state $\Psi_{\mathbf q}[s_2]$ as the one obtained by gluing regular simplices. 
\begin{equation}
\Psi_{\mathbf q}[s_2] = 
C_2 \ e^{
- (\alpha_2)_{ll'}{(j_l- j_l^{\scriptscriptstyle (2)})(j_{l'}- j_{l'}^{\scriptscriptstyle (2)})}
+i \Phi_l^{\scriptscriptstyle (2)}j_{l}}. 
\label{vuoto4}
\end{equation} 

Notice that something new happens with this amplitude, which did not happen in the previous cases: In general, the amplitude is suppressed unless the first order expansion of the Regge action matches the phases of the boundary state.  This gives a certain number of conditions on the internal spins (since these determine the dihedral angle appearing in the Regge action). Now, in the previous cases these conditions where sufficient to determine (the value around to which to expand) the internal spins uniquely. But this is not anymore true in the present case. Indeed, there are four external triangles, and therefore four conditions for the cancellation of the four phases $\Phi_l$ in (\ref{vuoto4}), while there are \emph{six} internal faces.  Therefore we can expect a two-parameter set of internal spins, whose contribution to the amplitude is \emph{not} suppressed by boundary-phase cancellation. 

On the other hand, there are six \emph{internal} deficit angles that appear in the expansion of the Regge action around the boundary $j_l$ and the internal references 
$j^{(2)}_{IJ}$, with $I<J$ and $I,J=2,3,4,5$, where now $j_l=(j^{\rm s}_{1v})$, with $v=1,4$ and $I,J,K,L=2,3,4,5$, $I<J$, $K<L$ 
\begin{eqnarray}
S_{Regge}({j^{\rm s}_n,j_{IJ}}) = \tilde \phi^{(2)}_n j^{\rm s}_{1v} +  \frac{1}{2} G_{ll'} \, \delta j_l \delta j_{l'}+\tilde \phi^{(2)}_n j^{\rm s}_{1v} + (\tilde \phi^{(2)u}_{I J}+\tilde \phi^{(2)d}_{I J}) j_{I J} + \\ \nonumber
+ G_{(IJ)l'} \, \delta j_{IJ} \delta j_{l'}+ \frac{1}{2} G_{(IJ),(KL)} \, \delta j_{IJ} \delta j_{KL}, 
\end{eqnarray}
and the give unmatched phases
\begin{equation}
\sum_{I,J=2,3,4,5;I<J}\Big(\tilde \phi^{(2)u}_{IJ} + \tilde \phi^{(2)d}_{IJ}\Big)j^{2}_{IJ}. 
\end{equation}
Thus, it appears that this terms is suppressed as well.  However, we expect this term to have a divergence, due to the presence of a bubble; does the divergence show in this  large $j$ limit?

\subsection{Subleading corrections}

Remarkably, none of the $\lambda^2$ terms considered here appear to give just a second order corrections in $1/j$ to the propagator.  The first case we have considered, namely the $4\to 1\to 4$ case, which is not suppressed, gives a contribution at the  
leading $1/j$ order. (This would happens for the other terms as well if we disregard the exponential suppression factor.)  Notice that in this case there is no first order term for a boundary state that has only support on the graph $\Gamma_8$.

\begin{figure}[h]
\begin{center}
\includegraphics[height=3.7cm]{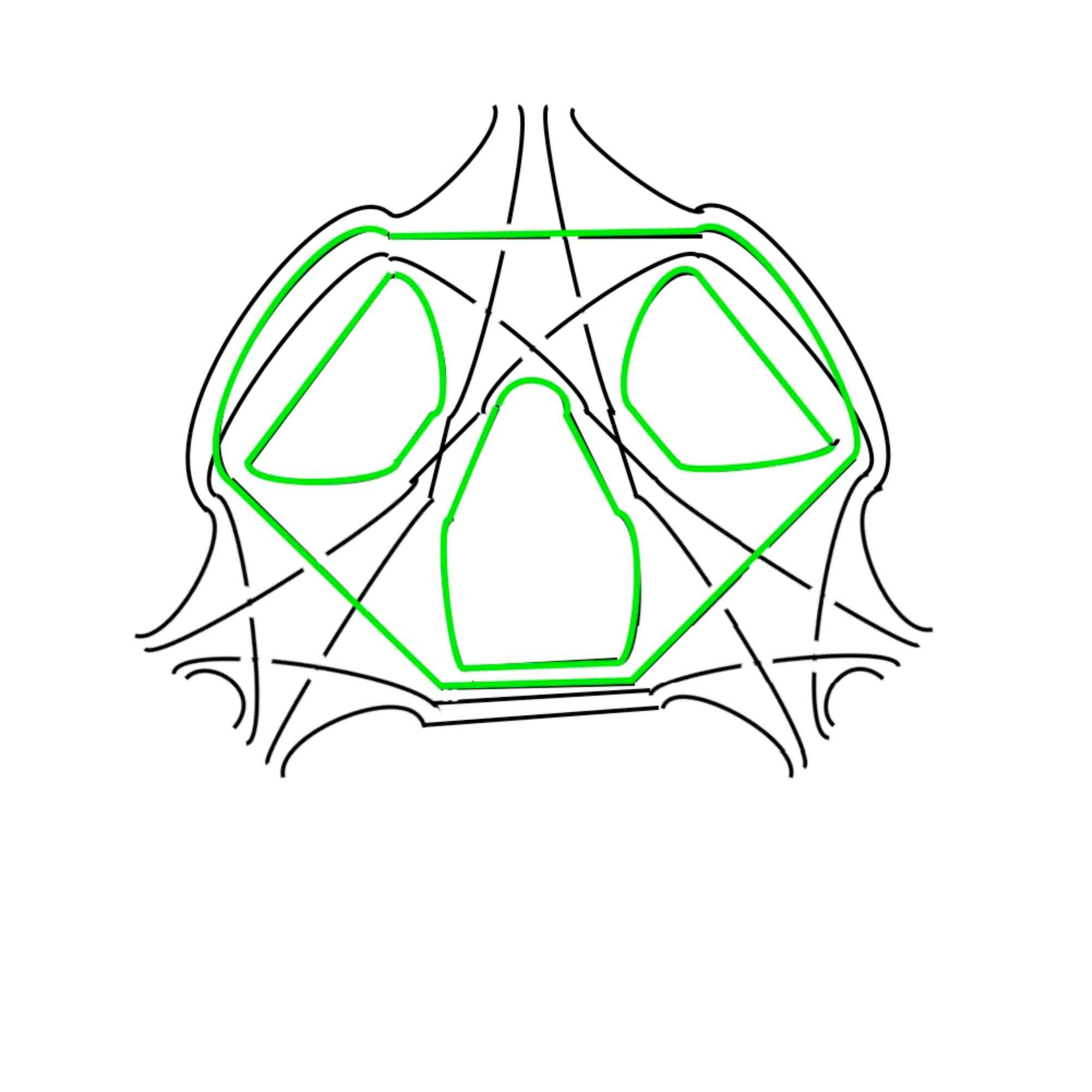}
\includegraphics[height=3.7cm]{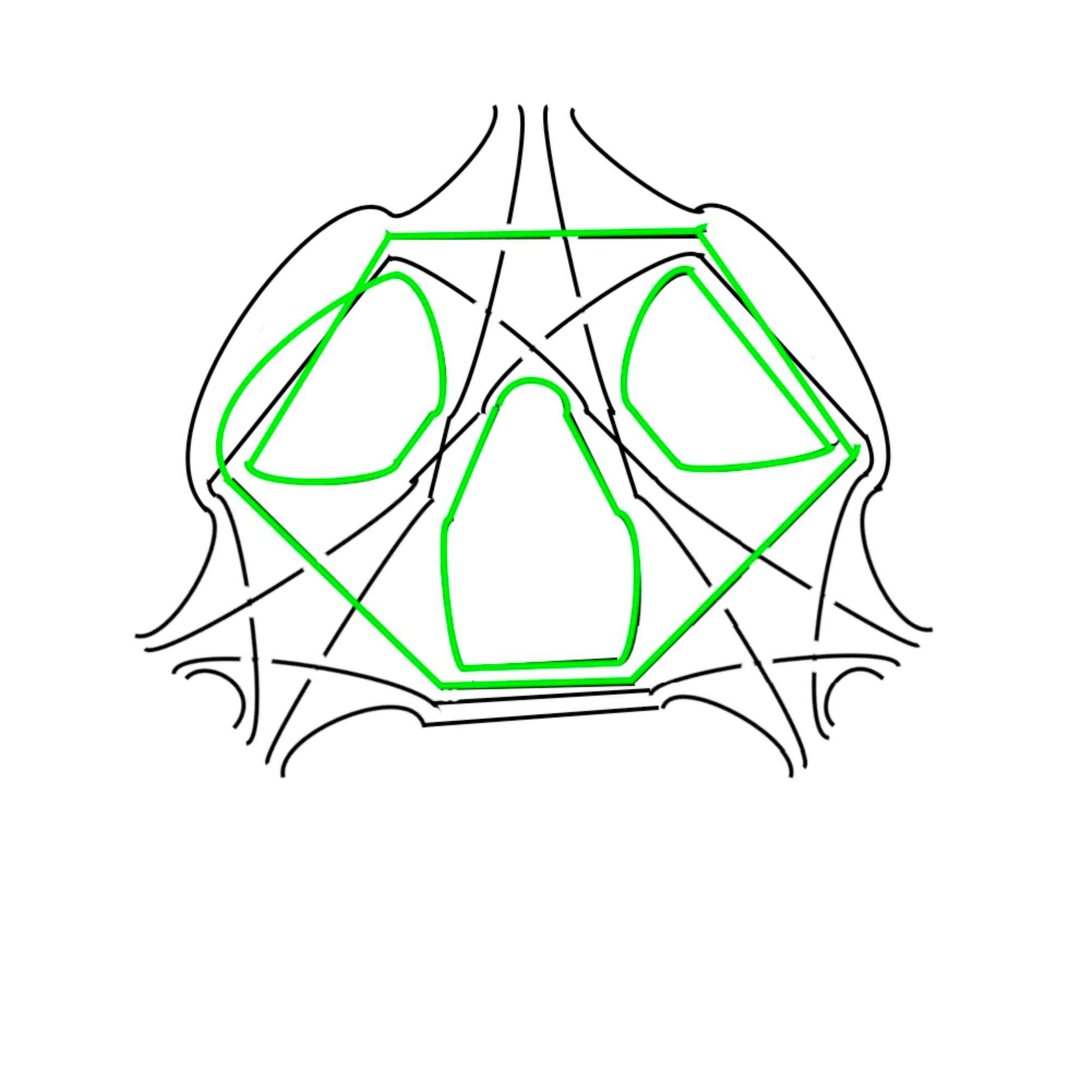}
\includegraphics[height=3.7cm]{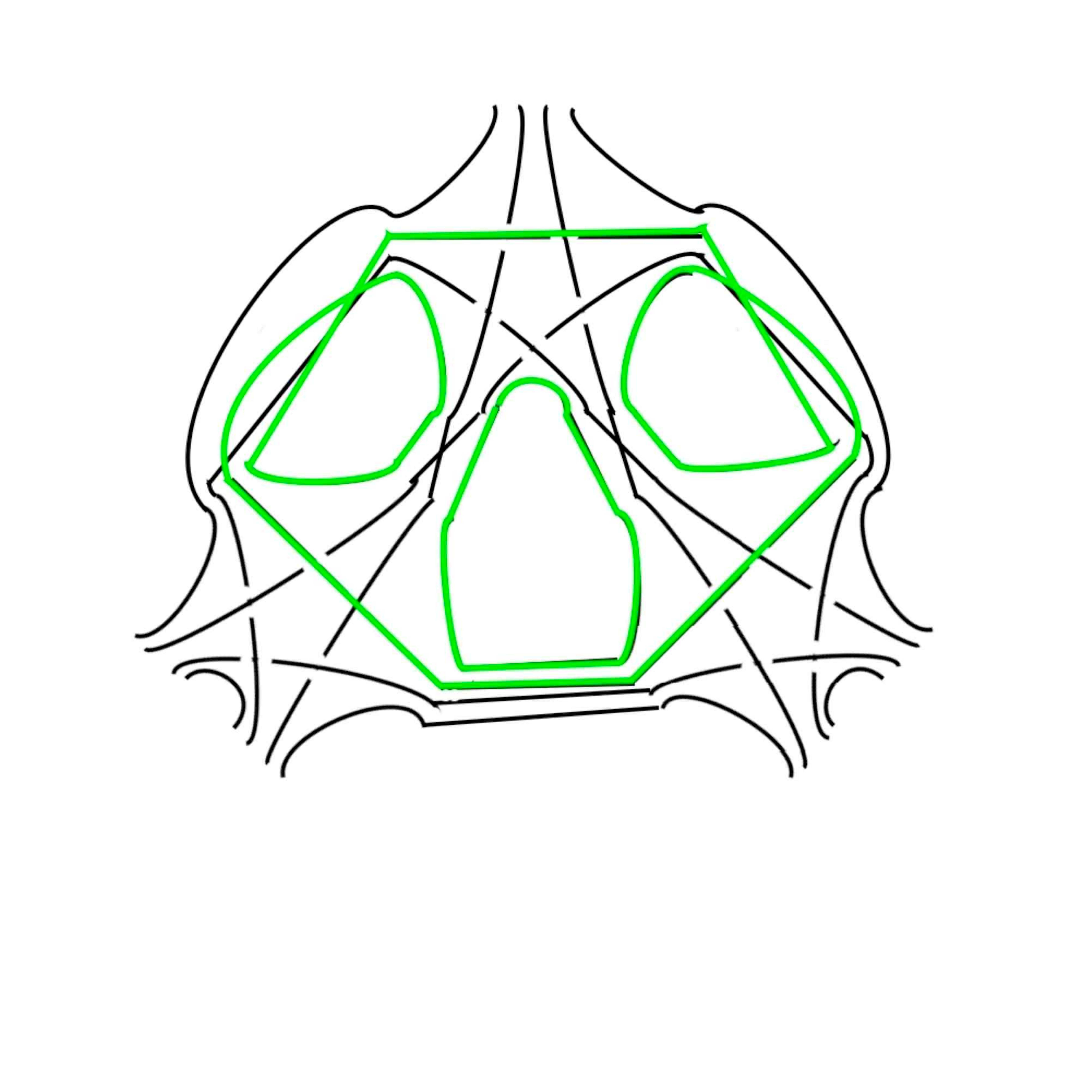}
\includegraphics[height=3.7cm]{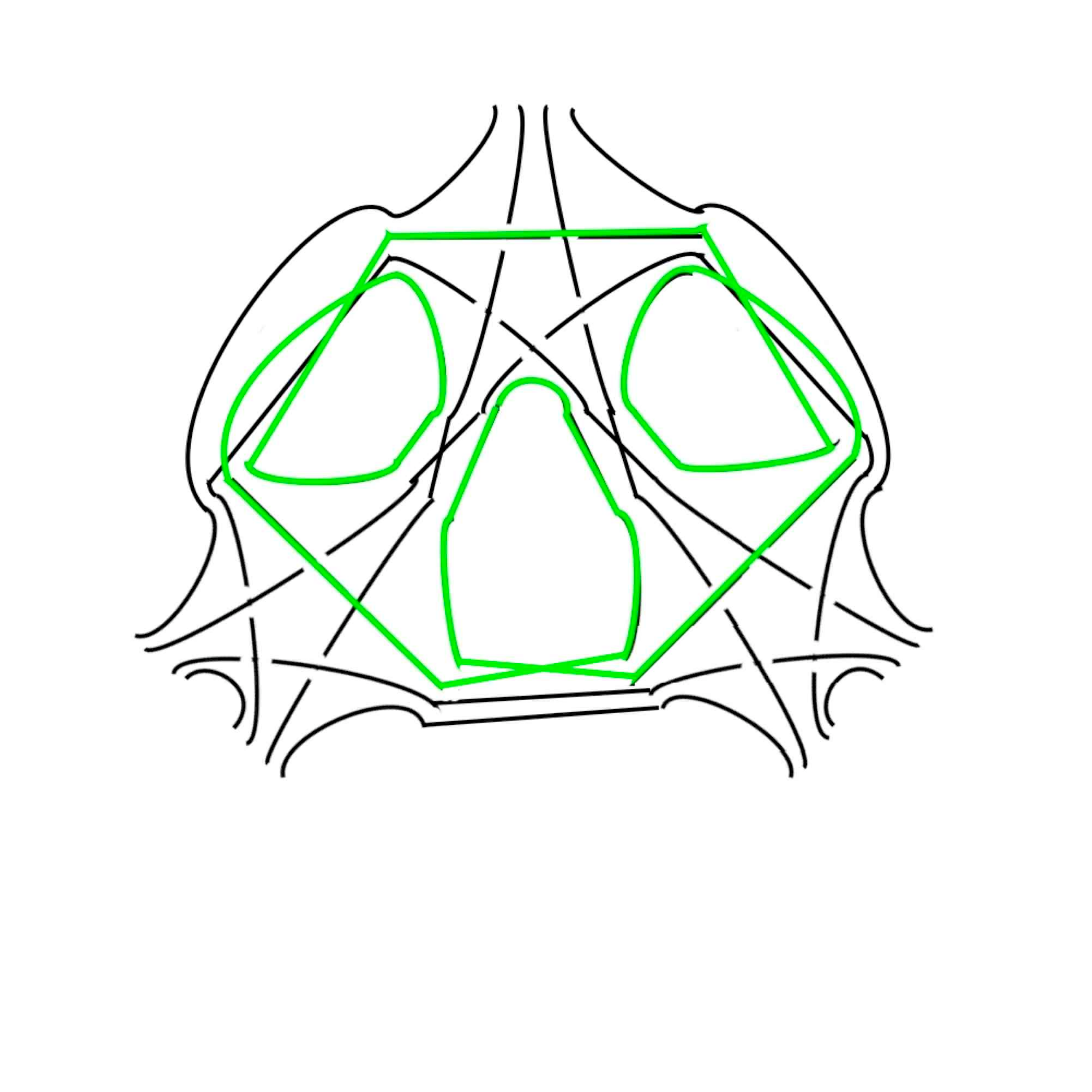}
\vskip-1cm
\caption{Different permutations in the propagators give a different number of faces: an example.}
\end{center}
\end{figure}

Of course terms of higher orders in $1/j$ abound as contributions to the amplitude.  First, we have expanded the vertex amplitude in $1/j$: all the vertex subleading terms will contribute to the amplitude.  More interestingly, recall that the propagator of the GFT includes the sum over permutations (\ref{s}). We have systematically considered only the term with the highest power in $j$ of this sum.  The other terms give lower powers in $j$.   

Notice in fact that the same sum over permutations appear in the normalization of the amplitude which is fixed by (\ref{norm_LQG}). The dominant term of the normalization will always be the one with highest power in $j$, and therefore the other terms in the amplitude will contribute as increasing powers in $1/j$.  It is tempting to speculate that these terms  are probably to be interpreted  as related to the corrections of the Newton potential, in agreement with standard correction obtained with quantum-field-theoretical techniques discussed in the literature (see \cite{liu,dono})
\begin{equation}
\label{potenzialecorretto}
V(r)=a_1\frac{Gm_1m_2}{r} \Big(1+a_2\frac{G(m_1+m_2)}{c^2}\frac{1}{r}+a_3\Big(\frac{G\hbar}{c^3}\Big)^2\frac{1}{r^2}\Big) + ... 
\end{equation}
with $a_1$, $a_2$ and $a_3$ numerical coefficients.  The actual calculation of these coefficients using the techniques developed here, however, requires more work, of which the one presented here is only a first step. 

\section{Conclusions}

We have explored some second-order contributions to the loop quantum gravity scattering amplitudes.  Our results are preliminary, and a more extensive study of these terms is needed.  Some general considerations appear nevertheless to be possible, and some interesting phenomena have appeared. 

\begin{enumerate}
\item Second order terms do not appear to spoil the correct large distance behavior of the two-point function.

\item The dominant contribution for the terms considered here appear to come from the two-complex with the maximal number of faces, which minimizes the complexity of the topology of its dual triangulation.  In other words, triangulations with very funny topologies appear to contribute less, at large distance.  

\item Terms of higher order in $\lambda$ contribute to the dominant $1/j$ term of the propagator, a result perhaps unexpected, but that was already pointed out in \cite{carlo2}. (This is the result of the $1\to 4 \to 1$ case.) 

\item The amplitude is suppressed unless the triangulation admits configurations where internal deficit angles appropriately vanish.   Thus, triangulations that admit only geometries that cannot solve the (discretized) Einstein equations do not contribute in the large $j$ limit. (This is the result of the $2\to 3 \to 2$ case.) 

\item Apparently, however, only triangulations that admit a flat geometry seem not to be suppressed.  This appears to be a problem, since it is \emph{Ricci} flatness, which seems the physically reasonable requirement.  This might be a problem of the model that we have used, and we think it should be clarified in the context of the new models \cite{pereira,Freidel:2007py,flipped,engle}.  (This is the result of the $3\to 2 \to 3$ case.) 

\item When the bulk of the triangulation is sufficiently more complex than the boundary, the internal spins are not fixed by the boundary geometry. The consequences of this case on the semiclassical limit are not yet clear to us. (This is the result of the $4\to 1 \to 4$ case.) 

\item Finally, it is not clear to us what happens in the large $j$ limit to the bubble divergence which is  expected in the $4\to 1 \to 4$ case. 

\end{enumerate}

Many other aspects of the problem remain unclear.  Among the most important, are how to work with general boundary states with components on different graphs, and to understand which one is the physical regime where the expansion in powers of $\lambda$ is viable. We think that continuing a concrete systematical exploration of the amplitudes may be useful path for addressing these questions. 

\vskip1cm

Thanks to Eugenio Bianchi for numerous useful comments.

The work of D.M. was partially supported by Fondazione Della Riccia and by EGIDE program of the French embassy in Italy.

\end{document}